\documentclass[11pt]{article}
\usepackage{amssymb}
\usepackage{amsmath}
\usepackage[dvips]{graphicx}
\usepackage{slashed}
\usepackage{hyperref}

\def\baselinestretch{1.4}

\begin{document}

\def\eg{{\it e.g.}}
\newcommand{\nc}{\newcommand}
\nc{\rnc}{\renewcommand}
\rnc{\d}{\mathrm{d}}
\nc{\D}{\partial}
\nc{\K}{\kappa}
\nc{\bK}{\bar{\K}}
\nc{\bN}{\bar{N}}
\nc{\bq}{\bar{q}}
\nc{\vbq}{\vec{\bar{q}}}
\nc{\g}{\gamma}
\nc{\lrarrow}{\leftrightarrow}
\nc{\rg}{\sqrt{g}}
\rnc{\[}{\begin{equation}}
\rnc{\]}{\end{equation}}
\nc{\bea}{\begin{eqnarray}}
\nc{\eea}{\end{eqnarray}}
\nc{\nn}{\nonumber}
\rnc{\(}{\left(}
\rnc{\)}{\right)}
\nc{\q}{\vec{q}}
\nc{\x}{\vec{x}}
\rnc{\a}{\hat{a}}
\nc{\ep}{\epsilon}
\nc{\tto}{\rightarrow}
\rnc{\inf}{\infty}
\rnc{\Re}{\mathrm{Re}}
\rnc{\Im}{\mathrm{Im}}
\nc{\z}{\zeta}
\nc{\mA}{\mathcal{A}}
\nc{\mB}{\mathcal{B}}
\nc{\mC}{\mathcal{C}}
\nc{\mD}{\mathcal{D}}
\rnc{\H}{\mathcal{H}}
\rnc{\L}{\mathcal{L}}
\nc{\<}{\langle}
\rnc{\>}{\rangle}
\nc{\fnl}{f_{NL}}
\nc{\gnl}{g_{NL}}
\nc{\fnleq}{f_{NL}^{equil.}}
\nc{\fnlloc}{f_{NL}^{local}}
\nc{\vphi}{\varphi}
\nc{\Lie}{\pounds}
\nc{\half}{\frac{1}{2}}
\nc{\bOmega}{\bar{\Omega}}
\nc{\bLambda}{\bar{\Lambda}}
\nc{\dN}{\delta N}
\nc{\gYM}{g_{\mathrm{YM}}}
\nc{\geff}{g_{\mathrm{eff}}}

\begin{titlepage}

\begin{center}

\hfill {\small ITFA-10-23}

\vskip 2 cm {\LARGE \bf Holographic Non-Gaussianity} 
\vskip 1.1 cm {\bf\large Paul McFadden${}^1$ and Kostas Skenderis$\,{}^{1,2,3}$}
\\ {\vskip 0.5cm \it\small
${}^1\,$Institute for Theoretical Physics,
${}^2\,$Gravitation and Astro-Particle Physics Amsterdam, \\
${}^3\,$Korteweg-de$\,$Vries Institute for Mathematics, \\
Science Park 904, 1090 GL Amsterdam, the Netherlands.}

{\vskip 0.2cm \small
{\it E-mail:} {\tt P.L.McFadden@uva.nl, K.Skenderis@uva.nl} }

\end{center}

\vskip 1 cm

\begin{abstract}

We investigate the non-Gaussianity of primordial cosmological perturbations within our recently proposed holographic description of inflationary universes. We derive a holographic formula that determines the bispectrum of cosmological curvature perturbations in terms of correlation functions of a holographically dual three-dimensional non-gravitational quantum field theory (QFT). This allows us to compute the primordial bispectrum for a universe which started in a non-geometric holographic phase, using perturbative QFT calculations. Strikingly, for a class of models
specified by a three-dimensional super-renormalisable QFT, the primordial bispectrum is of exactly the factorisable equilateral form with $\fnleq=5/36$, irrespective of the details of the dual QFT. 

A by-product of this investigation is a holographic formula for the three-point function  
of the trace of the stress-energy tensor along general holographic RG flows, which should have applications outside the remit of this work.

\end{abstract}

\end{titlepage}

\thispagestyle{empty}
\tableofcontents

\newpage
\setcounter{page}{1}

\section{Introduction}

Primordial cosmological perturbations and their properties provide some of the best observational clues to the
physics of the very early universe.
Acting as the seed for structure formation, these initial inhomogeneities left behind an imprint in the cosmic microwave background (CMB), and in the distribution of large-scale structure, through which their properties may be directly inferred.
To date, there is no compelling evidence for any departure from Gaussianity: the different Fourier modes of the perturbations appear to be uncorrelated, with random phases.  This implies that all higher correlation functions of the primordial perturbations may be expressed in terms of the 2-point function, or equivalently its Fourier transform, the power spectrum $\Delta_S^2(q)$.  In particular, the 3-point function and all odd higher-order correlators should vanish.

Nevertheless, the significant improvement in observational data expected in the very near future may change
this situation.
Quantitatively, the Fourier transform of the 3-point function of curvature perturbations, the scalar bispectrum, may be parameterised by an overall amplitude, $\fnl$, along with a momentum dependence or `shape' function (see
\cite{Bartolo:2004if, Koyama:2010xj, Chen:2010xk, Komatsu:2010hc} for reviews).
From the WMAP $7$-year data \cite{Komatsu:2010fb}, the observational constraints on $\fnl$, 
for two specific choices of shape function, are
\[
 \fnlloc = 32 \pm 21, \qquad \fnleq = 26\pm140, 
\]
where the first value corresponds to 
the `local' shape \cite{Gangui:1993tt, Verde:1999ij, Komatsu:2001rj} and the second corresponds to 
the `equilateral' shape \cite{Creminelli:2005hu}.
In just a few years' time, the results from the Planck satellite 
are expected to further reduce the uncertainty on $\fnl$ to approximately  $\Delta \fnl \sim 5$ \cite{Komatsu:2001rj}.
Constraints deriving from observations of large-scale structure may also in future be competitive with those from the CMB \cite{Desjacques:2010nn}.

In view of its power to elucidate features of
the mechanism through which the primordial perturbations were generated,
any future detection of primordial non-Gaussianity will be of paramount importance for cosmology.
In the context of inflation, the leading candidate for such a mechanism,
primordial non-Gaussianity reveals details of
the dynamical interactions present during the inflationary epoch. 

In the simplest models of inflation, based on a single scalar field slowly rolling down a potential, these interactions are suppressed by powers of the slow-roll parameters giving rise to an $\fnl$ of first order in slow-roll ({\it i.e.}, $\fnl \sim O(0.01)$) \cite{Gangui:1993tt, Salopek:1990jq, Falk:1992sf,  Acquaviva:2002ud, Maldacena:2002vr}.
More elaborate inflationary models including, {\it e.g.}, multiple scalar fields \cite{Linde:1996gt, Gordon:2000hv, Lyth:2002my, Bernardeau:2002jy, Seery:2005gb}, non-canonical kinetic terms \cite{Alishahiha:2004eh, Seery:2005wm, Chen:2006nt}, inhomogeneous reheating \cite{Dvali:2003ar}, features in the potential \cite{Chen:2006xjb,Chen:2008wn,Flauger:2010ja}, or initial state modifications \cite{Holman:2007na, Meerburg:2009fi, Meerburg:2009ys} all give rise to a considerable range of $\fnl$ values, as well as different predictions for the shape function.
Since there is often little to distinguish models at the level of the power spectrum, non-Gaussianity provides a powerful means of
observationally discriminating between the various inflationary candidates,
as well as serving to constrain other non-inflationary scenarios \cite{Lehners:2010fy}.
Note for these purposes it is necessary to distinguish the {\it primordial} non-Gaussianity generated during the inflationary epoch
from that generated in later stages of cosmological evolution ({\it e.g.}, by the nonlinear evolution of perturbations after they re-enter the horizon
in the matter and radiation eras, and by nonlinearities in the relation between metric fluctuations and temperature fluctuations in the CMB).
With a sufficiently accurate understanding of their physics \cite{Pitrou:2010sn, Bartolo:2010qu}, these latter sources of non-Gaussianity may be subtracted off to leave the primordial component of principal interest for constraining cosmological models.

\begin{figure}
\begin{center}
\includegraphics[width=0.63\textwidth]{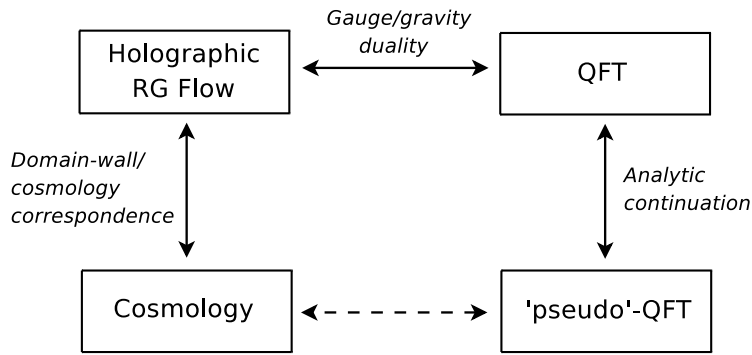}  
\end{center}
\caption{\label{holysq}
The `pseudo'-QFT dual to inflationary cosmology is operationally defined
using the domain-wall/cosmology correspondence and standard gauge/gravity duality.
}
\end{figure}

In the present paper, we initiate the investigation of
primordial non-Gaussianity for our recently proposed holographic description of inflationary cosmology \cite{McFadden:2009fg, McFadden:2010na, McFadden:2010jw}.
The key to this description is
the holographic framework depicted in Figure \ref{holysq}, which connects four-dimensional inflationary universes with
three-dimensional non-gravitational QFTs.
The basic ingredients are ordinary gauge/gravity duality (corresponding to the uppermost arrow in the figure), combined with the domain/wall cosmology correspondence \cite{Cvetic:1994ya,Skenderis:2006jq, Skenderis:2006fb} (lefthand vertical arrow), a simple analytic continuation relating cosmologies to domain-wall spacetimes describing holographic RG flows.
This bulk analytic continuation may be re-expressed in the language of the dual QFT (righthand vertical arrow), whereupon it takes
correlators of the dual QFT to correlators of the so-called `pseudo'-QFT, which we propose is dual to the original cosmology (lower dashed arrow).

On the basis of this framework, cosmological observables may be re-expressed in terms of correlators of the dual QFT.
At the level of linear perturbation theory, exact formulae have been derived
relating the cosmological scalar and tensor power spectra with the 2-point function for the stress-energy tensor of the dual QFT \cite{McFadden:2009fg,McFadden:2010na}.
The first major goal of the present work will be
to extend this correspondence to {\it quadratic order} in perturbation theory.
Focusing on the scalar bispectrum, the
cosmological observable most relevant to the 
present and forthcoming observational data, we will show how this quantity may be naturally re-expressed in terms of the 3-point function of the dual stress-energy tensor\footnote{For earlier work in a similar spirit, see \cite{Seery:2006tq}.}.
Analogous formulae for other non-Gaussian cosmological observables may be derived by similar methods and will be reported elsewhere \cite{McFadden:2011kk}.
The Hamiltonian holographic renormalisation method we develop for computing
3-point functions
is based on \cite{Papadimitriou:2004ap, Papadimitriou:2004rz}, and
may well be of utility in a wider holographic context\footnote{Earlier
work on the computation of 3-point functions for holographic RG flows may be found in
\cite{Bianchi:2003bd,Bianchi:2003ug}.}.

One striking feature of holographic dualities is that they are
strong/weak coupling dualities, meaning that when one description is
weakly coupled, the other is strongly coupled, and vice versa.  
In the regime where the dual QFT is strongly coupled then, the
gravitational description is weakly coupled and our holographic
formulae should (and indeed they do) reproduce the results of standard
single-field inflation.  In this situation, the application of our
holographic framework offers a fresh perspective, and may lead to new insights,
but offers no new predictions.

In the regime in which the dual QFT is weakly coupled, however, the corresponding gravitational description is instead {\it strongly coupled} at very early times.
We emphasize that by `strongly coupled' gravity we do {\it not} mean that the perturbative
fluctuations around the background FRW spacetime are strongly coupled, but rather, that the description in terms of metric fluctuations is itself not
valid. This is a non-geometric `stringy' phase.
A geometric description emerges only asymptotically, and at late times one recovers a specific accelerating FRW spacetime (to be matched to conventional hot big bang cosmology), along with a specific set of inhomogeneities.  Crucially, these inhomogeneities are not linked with a perturbative quantisation around the FRW spacetime as in conventional inflation, but rather, they originate from the dynamics of the dual weakly coupled QFT.
Holography thus suggests a natural generalisation of the inflationary mechanism to strongly coupled gravity, in which the properties of cosmological perturbations may be determined through three-dimensional perturbative QFT calculations.

In order to perform such calculations,
it is necessary to specify more precisely the nature of the dual QFT.
Ideally, one would be able to deduce this from first principles via some string/M-theoretic construction.
In the absence of such a construction, we will instead pursue a (holographic)
phenomenological approach. As with other known holographic dualities, the dual QFT will in general involve scalars,
fermions and gauge fields, and it should admit a large $N$ limit. The
question is then whether one can find a theory which is compatible
with current observations.

An additional guiding principle is to consider QFTs of the type that feature in the 
description of holographic RG flows.
This holographic description is well understood for two classes of domain-wall spacetimes,
namely, those that are asymptotically anti-de Sitter, and those with asymptotically power-law scaling.
Under the domain-wall/cosmology correspondence, these correspond respectively to asymptotically de Sitter, and to asymptotically power-law cosmologies.  The first class of domain-wall solutions describe QFTs that are either deformations of CFTs, or else CFTs in a nontrivial vacuum state, while the second class describes QFTs with a single dimensionful parameter in the regime in which the dimensionality of the coupling constant drives the dynamics \cite{Kanitscheider:2008kd}. Examples of such dualities are provided by considering the near-horizon
limit of the non-conformal branes \cite{Itzhaki:1998dd,Boonstra:1998mp}.
The detailed holographic dictionary for these theories has been
worked out only relatively recently \cite{Wiseman:2008qa,Kanitscheider:2008kd,Kanitscheider:2009as}.
These theories are characterised by the fact that they have a `generalised conformal structure' \cite{Jevicki:1998qs,Jevicki:1998ub,Jevicki:1998yr,Kanitscheider:2008kd}. In particular, all
terms in the Lagrangian have the same scaling dimension, which is however different from the
spacetime dimension.

Focusing on this second class, we will consider here
super-renormalisable theories that contain one
dimensionful coupling constant.  A prototype example is three-dimensional $SU(N)$
Yang-Mills theory coupled to a number of scalars and fermions, all transforming in the adjoint of $SU(N)$.
Theories of this type are typical in AdS/CFT
where they appear as the worldvolume theories of D-branes.
A general such model that admits a large $N$ limit is
\[
\label{Lfree}
S = \frac{1}{g_{\mathrm{YM}}^2}\int \d^3 x\,
 \mathrm{tr}\left( \frac{1}{2} F^I_{ij}F^I_{ij} +  \frac{1}{2} (\D\phi^J)^2
+  \frac{1}{2} (\D\chi^K)^2
+ \bar{\psi}^L \slashed{\D} \psi^L + \mathrm{interactions}\, \right),
\]
where we consider $\mathcal{N}_A$ gauge fields $A^I$  ($I=1,\ \ldots,\ \mathcal{N}_A)$,
$\mathcal{N}_\phi$ minimal scalars $\phi^J$ ($J=1,\ \ldots,\ \mathcal{N}_\phi)$,
$\mathcal{N}_\chi$ conformal scalars $\chi^K$ ($K=1,\ \ldots,\ \mathcal{N}_\chi)$
 and $\mathcal{N}_\psi$ fermions $\psi^L$  ($L=1,\ \ldots,\  \mathcal{N}_\psi)$.
Note that $g_{\mathrm{YM}}^2$ has dimension one in three dimensions.
In general, the Lagrangian \eqref{Lfree} will also contain dimension-four interaction terms (see \cite{McFadden:2010na}).
We will leave these interaction unspecified, however, as they do not contribute to the leading order calculations we will perform here.

As shown in \cite{McFadden:2009fg, McFadden:2010na, McFadden:2010jw}, it is straightforward to find holographic models of this form that yield cosmological predictions compatible with current observations.
Generically, we obtain a nearly scale-invariant spectrum of small amplitude perturbations, where the overall amplitudes of the power spectra scale as $\sim 1/N^2$, and their deviation from scale invariance is of order the dimensionless effective coupling, $g_{\mathrm{eff}}^2 = g_{\mathrm{YM}}^2 N/q$, where $q$ is a typical momentum.  In particular, the small observed amplitude $\sim O(10^{-9})$ of the scalar power spectrum implies $N\sim O(10^4)$ consistent with the large $N$ limit, while the smallness of the observed deviation from scale invariance $\sim O(10^{-2})$ is consistent with the assumed weak coupling limit where $g_{\mathrm{eff}}^2 \ll 1$.
Through appropriate choice of the field content of the dual QFT, it is further possible to satisfy the current observational upper bounds on the ratio of tensors to scalars.

A distinctive prediction of these holographic models is a well-defined running of the spectral indices:
in the scalar case, for example, the running is given by minus the deviation from scale invariance.  This prediction is markedly different from conventional slow-roll inflation, where the running is heavily suppressed, and may potentially be excluded by the forthcoming Planck data \cite{McFadden:2010jw}.

Having established a precise holographic formula linking the cosmological bispectrum to the 3-point function for the dual stress-energy tensor, our second major goal will be to use this formula to predict the cosmological non-Gaussianity arising from a weakly coupled dual QFT of the form \eqref{Lfree}.  These predictions will complement those for the power spectra discussed above, and may potentially reveal further distinctive observational signatures of holographic models.

The remainder of this paper is organised as follows.  In Section \ref{DWC_section}, we discuss perturbation theory for domain-walls and cosmologies at quadratic order: after defining the metric fluctuations and the gauge-invariant curvature perturbation $\z$, we evaluate the cubic interaction Hamiltonian and set up the domain-wall/cosmology correspondence.  We also introduce  response functions relating the curvature perturbation to its corresponding canonical momentum; these response functions will play a central role in our subsequent holographic analysis.  In Section \ref{Bispec_section}, we summarise the calculation of the cosmological bispectrum and show how to re-write it in terms of response functions.  We then proceed, in Section \ref{Hol_section}, with the holographic calculation of the 3-point function for the trace of the dual stress-energy tensor.  After a brief introduction to the radial Hamiltonian holographic renormalisation methods we use, we compute the holographic 3-point function for both asymptotically AdS and asymptotically power-law domain-walls.  Finally, in Section \ref{Hol_formulae}, we combine these results to show how the cosmological observables may be expressed in terms of correlation functions of the dual QFT,
and in Section \ref{HolNG_section},  after performing the relevant QFT calculations, we arrive at a holographic prediction for primordial non-Gaussianity.

\section{Perturbed domain-walls and cosmologies}
\label{DWC_section}

\subsection{Defining the perturbations}

Domain-walls and cosmologies may be described in a unified fashion via the ADM metric
\[
 \d s^2 = \sigma N^2 \d z^2 + g_{ij}(\d x^i+N^i\d z)(\d x^j+N^j\d z),
\]
where the perturbed lapse and shift functions may be written to second order as
\[
\label{gij_def}
 N=1+\dN(z,\vec{x}), \qquad N_i=g_{ij}N^j=\dN_i(z,\vec{x}), \qquad g_{ij}= a^2(z)(\delta_{ij}+h_{ij}(z,\vec{x})),
\]
with $\sigma=+1$ for a Euclidean domain-wall\footnote{\renewcommand{\baselinestretch}{1}\normalsize\small
A Lorentzian domain-wall can be obtained by
continuing one of the $x^i$ coordinates to become time \cite{Skenderis:2006fb}. The
continuation to a Euclidean domain-wall is convenient, however, since the QFT vacuum implicit in the
Euclidean formulation maps to the Bunch-Davies vacuum on the cosmology side. Other choices of vacua
may be accommodated using the real-time formalism of \cite{Skenderis:2008dh}. This
is an interesting extension that we leave for future work.}
(whereupon $z$ becomes the transverse radial coordinate) and $\sigma=-1$ for a cosmology (whereupon $z$ becomes the cosmological proper time).  The spatial indices $i,j$ run from $1$ to $3$, and we have assumed (for simplicity) the background geometry to be spatially flat.

The $\delta g_{00}$ metric perturbation is then
\[
 \delta g_{00} = 2\sigma \phi = \sigma (2\dN+\dN^2)+a^{-2} \dN_i \dN_i,
\]
where here, and in the remainder of the paper, we adopt the convention that repeated covariant indices are summed using the Kronecker delta 
(in contrast, an index is raised or lowered by the full metric).
The remaining perturbations may be decomposed into scalar, vector and tensor pieces according to
\[
\label{metric_perts}
 \dN_i = a^2(\nu_{,i} + \nu_i), \qquad h_{ij} = -2\psi\delta_{ij}+2\chi_{,ij}+2\omega_{(i,j)}+\g_{ij},
\]
where the vector perturbations $\nu_i$ and $\omega_i$ are transverse, and the tensor perturbation $\g_{ij}$ is transverse traceless.
We similarly decompose the inflaton $\Phi$ into a background piece $\vphi$ and a perturbation $\delta\vphi$,
\[
\label{fullphi_def}
 \Phi(z,\vec{x}) = \vphi(z)+\delta\vphi(z,\vec{x}).
\]
These formulae are understood to hold to second order in perturbation theory.

We define $\z(z,\vec{x})$, the curvature perturbation on uniform energy density slices, so that in comoving gauge where $\delta\vphi$ vanishes, the spatial part of the perturbed metric reads
\[
g_{ij} = a^2 e^{2\z}[e^{\hat{\g}}]_{ij} =a^2 e^{2\z}(\delta_{ij}+\hat{\g}_{ij}+\half \hat{\g}_{ik}\hat{\g}_{kj}),
\]
where $\hat{\g}_{ij}$ is transverse traceless\footnote{\renewcommand{\baselinestretch}{1}\normalsize\small
We have chosen this definition so as to coincide with most of the recent literature on non-Gaussian perturbations, in particular \cite{Maldacena:2002vr}.  Note that in our previous articles \cite{McFadden:2009fg, McFadden:2010na} we defined $\z$ at linear order to be instead the comoving curvature perturbation, which differs by a sign.}.
 This definition may then be straight-forwardly recast  into the general gauge-invariant form (see Appendix \ref{App_GT} for details)
\begin{align}
\label{zeta_gi}
\z &= -\psi-\frac{H}{\dot{\vphi}}\delta\vphi -\psi^2
+\Big(\dot{H}-\frac{H\ddot{\vphi}}{\dot{\vphi}}\Big)\frac{\delta\vphi^2}{2\dot{\vphi}^2}
+\frac{H}{\dot{\vphi}^2}\delta\vphi\delta\dot{\vphi}+\frac{H}{\dot{\vphi}}\hat{\xi}_k\delta\vphi_{,k} \nn \\
&\quad  +\frac{1}{4}\pi_{ij}\Big(\frac{\sigma}{a^2\dot{\vphi}^2}\delta\vphi_{,i}\delta\vphi_{,j}
-\frac{2}{a^2 \dot{\vphi}}\dN_i\delta\vphi_{,j}
-\frac{\delta\vphi}{\dot{\vphi}}\dot{h}_{ij} - 2\hat{\xi}_{k,i}h_{jk} - \hat{\xi}_k h_{ij,k}  \nn \\
& \qquad\qquad\qquad 
+\hat{\xi}_{k,i}\hat{\xi}_{k,j} +2\psi\g_{ij} -\half\g_{ik}\g_{kj}\Big),
\end{align}
where $\hat{\xi}_k=\chi_{,k}+\omega_k$.  Here, and throughout, we use dots to denote differentiation with respect to $z$ and we set $H=\dot{a}/a$.
The transverse projection operator $\pi_{ij}$ is defined as
\[
 \pi_{ij} = \delta_{ij}-\frac{\D_i\D_j}{\D^2}.
\]

The physical significance of $\z$ is that it is conserved on super-horizon scales, in the absence of entropy perturbations.
This holds to all orders in perturbation theory \cite{Langlois:2005qp}, and serves to connect the behaviour of modes as they exit the horizon during the inflationary epoch to their initial conditions at horizon re-entry in the subsequent radiation- and matter-dominated eras.

\subsection{Equations of motion} 

In the ADM formalism, the combined domain-wall/cosmology action for a single minimally coupled scalar field takes the form
\[
S= \frac{1}{2\K^2}\int\d^4x N\sqrt{g} \left[K_{ij}K^{ij}-K^2+N^{-2}(\dot{\Phi}-N^i\Phi_{,i})^2
+\sigma \(-R+g^{ij}\Phi_{,i}\Phi_{,j}+2\K^2 V(\Phi)\)\right], \nn
\]
where $\kappa^2=8\pi G$, $K_{ij}=[(1/2)\dot{g}_{ij}-\nabla_{(i}N_{j)}]/N$ is the extrinsic curvature of constant-$z$ slices,
and we have taken the scalar field $\Phi$ to be dimensionless.  In this expression, the spatial gradient and potential terms appear with positive sign for Euclidean domain-walls and with negative sign for Lorentzian cosmologies, as indeed they should.

We will restrict our consideration to background solutions in which the evolution of the scalar field $\vphi(z)$ is
(piece-wise) monotonic in $z$. 
For such solutions, $\vphi(z)$ can in principle be inverted to $z(\vphi)$, allowing the Hubble rate $H=\dot{a}/a$ to be re-expressed as a function of $\vphi$, {\it i.e.}, $H(z) = -(1/2)W(\vphi)$.  The complete equations of motion for the background then take the simple form
\[
\label{bgd_eom}
\frac{\dot{a}}{a} =-\frac{1}{2}W, \qquad \dot{\vphi}= W_{,\vphi}, \qquad 2\sigma\K^2V = (W_{,\vphi})^2-\frac{3}{2}W^2.
\]
In cosmology, this first-order formalism dates back to the work of \cite{Salopek:1990jq}, where it was obtained
by application of the Hamilton-Jacobi method.  For domain-walls, this formalism has been discussed from
variety of standpoints (gravitational stability, Hamilton-Jacobi method, fake supersymmetry) in \cite{Skenderis:2006jq, Skenderis:1999mm,DeWolfe:1999cp,deBoer:1999xf,Freedman:2003ax}. In this context,
the function $W(\vphi)$ is the `fake superpotential' ({\it i.e.}, when the domain-wall solution is a supersymmetric
solution of a supergravity theory, $W(\vphi)$ is the true superpotential).

Turning now to the perturbations, following Maldacena \cite{Maldacena:2002vr}, the cubic action for $\z$ may be derived by solving the Hamiltonian and momentum constraints and backsubstituting into the Lagrangian.  Keeping careful track of the sign $\sigma$, we find
\begin{align}
\label{Lagrangian}
S = \int\d^4x\,\L= \frac{1}{\K^2} \int\d^4x\,\Big[ a^3\ep \dot{\z}^2+\sigma a\ep(\D\z)^2  & - \frac{a^3\ep}{H}\dot{\z}^3+3a^3\ep\z\dot{\z}^2+\sigma a\ep\z(\D\z)^2
-2a^3\z_{,k}\hat{\nu}_{,k}\D^2\hat{\nu} \nn \\
& -\frac{a^3}{2}\big(\frac{\dot{\z}}{H}-3\z\big) \(\hat{\nu}_{,ij}\hat{\nu}_{,ij}-\D^2 \hat{\nu} \D^2\hat{\nu}\)\Big],
\end{align}
where $\ep=-\dot{H}/H^2=\dot{\vphi}^2/2H^2$ (note we do not use the slow roll approximation) and $\hat{\nu} = \ep\, \D^{-2}\dot{\z}+(\sigma/a^2 H)\z$.

In the Hamiltonian formalism, one then has the quadratic free Hamiltonian
\[
\label{H2}
H^{(2)} = \frac{1}{\K^2}\int\d^3\x \left[\frac{1}{4a^3\ep}\Pi^2-\sigma a \ep(\D\z)^2\right],
\]
where
\[
\Pi = \frac{\D(\K^2 \L)}{\D\dot{\z}}
\]
is ($\K^2$ times) the canonical momentum conjugate to $\z$, and the cubic interaction Hamiltonian
\begin{align}
\label{H3}
 H^{(3)} =-\int\d^3\x \,\L^{(3)} = \frac{1}{\K^2}\int\d^3\x\Big[ \frac{1}{8a^6\ep^2 H}&\Pi^3  -\frac{3}{4a^3\ep}\Pi^2\z - \sigma a\ep \z(\D\z)^2 + 2a^3 \z_{,k} \hat{\nu}_{,k} \D^2 \hat{\nu} \nn \\
& + \big(\frac{1}{4\ep H}\Pi-\frac{3a^3}{2}\z\big)(\hat{\nu}_{,ij}\hat{\nu}_{,ij}-\D^2\hat{\nu}\D^2\hat{\nu})\Big],
\end{align}
where in this expression $\hat{\nu} = (1/2a^3)\,\D^{-2}\Pi+(\sigma/a^2H)\z$.

Passing to momentum space, one finds
\bea
H^{(2)} &=& \frac{1}{\K^2}\int [\d q] \,\Big[ \frac{1}{4a^3\ep}\Pi(\q)\Pi(-\q)- \sigma a \ep q^2\,\z(\q)\z(-\q)\Big], \nn \\[2ex]
\label{H3decomp}
H^{(3)} &=& \frac{1}{\K^2}\int
[[\d q_1\d q_1\d q_3]] 
\Big[ \mA(q_i)\z(-\q_1)\z(-\q_2)\z(-\q_3)
+\mB(q_i)\Pi(-\q_1)\z(-\q_2)\z(-\q_3) \nn \\
&&\qquad\qquad\quad + \mC(q_i) \z(-\q_1)\Pi(-\q_2)\Pi(-\q_3) +\mD(q_i)\Pi(-\q_1)\Pi(-\q_2)\Pi(-\q_3)  \Big],
\eea
where here, and in the remainder of the paper, we will use the shorthand notations
\begin{align}
\label{shorthand}
&[\d q] \equiv \d^3 \vec{q} /(2\pi)^3, \qquad [[\d q_2 \d q_3]] \equiv (2\pi)^3\delta(\sum_i \q_i)[\d q_2][\d q_3], \nn\\
&\qquad  [[\d q_1\d q_2\d q_3]]\equiv (2\pi)^3\delta(\sum_i \q_i)[\d q_1][\d q_2][\d q_3].
\end{align}
The coefficients $\mA$, $\mB$, $\mC$ and $\mD$ may be written as 
\bea
\label{Adef}
\mA(q_i) &=& -\frac{1}{24 a H^2}\( 2P_{(4)}-P_{(2)}^2\)
 -\frac{\sigma a \ep}{6} P_{(2)},  \\[2ex]
\label{Bdef}
\mB(q_i) &=& \frac{1}{16a^4\ep H^3}\(2P_{(4)} -P_{(2)}^2\)
-\frac{\sigma}{8 a^2 H q_1^2}\(4q_1^4-2q_1^2P_{(2)}-2P_{(4)}+P_{(2)}^2 \),  \\[2ex]
\label{Cdef}
\mC(q_i) &=& -\frac{1}{32a^3\ep}\Big[24 + \frac{\ep}{q_2^2q_3^2}\(8q_1^4-4q_1^2P_{(2)}-2P_{(4)}+P_{(2)}^2\) \nn \\
&&\qquad\qquad\qquad +\frac{\sigma}{a^2H^2}\frac{1}{q_2^2q_3^2}\(P_{(2)}-q_1^2\)\(2P_{(4)}-P_{(2)}^2 \)\Big], \\[2ex]
\label{Ddef}
\mD(q_i) &=& \frac{1}{16a^6\ep H}\left[\frac{2}{\ep}-1+\frac{1}{12q_1^2q_2^2q_3^2}\(P_{(2)}^3-4P_{(2)}P_{(4)}+4P_{(6)} \)\right],
\eea
where the magnitudes $q_i = +\sqrt{{\vec{q}_i}^{\,\,2}}$ and the symmetric polynomials $P_{(n)}=\sum_i (q_i)^n$.

Writing $\mC_{213}=\mC(q_2,q_1,q_3)$, {\it etc.}, Hamilton's equations 
then read
\begin{align}
\label{H_eom1}
\dot{\z}(\q_1) &= (2\pi)^3\, \frac{\D(\K^2 H)}{\D\Pi(-\q_1)} \nn \\
&= \frac{1}{2a^3\ep}\Pi(\q_1)+\int [[\d q_2\d q_3]] \Big[\mB_{123}\z(-\q_2)\z(-\q_3)+2\mC_{213}\z(-\q_2)\Pi(-\q_3)  \nn \\
&\qquad\qquad\qquad\qquad\qquad\qquad\qquad\qquad\qquad +3\mD_{123}\Pi(-\q_2)\Pi(-\q_3)\Big], 
\end{align}
\begin{align}
\label{H_eom2}
\dot{\Pi}(\q_1) &= -(2\pi)^3 \frac{\D(\K^2 H)}{\D\z(-\q_1)} \nn \\
&= 2\sigma a \ep q_1^2 \z(\q_1) - \int[[\d q_2\d q_3]] \Big[3\mA_{123}\z(-\q_2)\z(-\q_3)
+2\mB_{213}\Pi(-\q_2)\z(-\q_3) \nn \\
&\qquad\qquad\qquad\qquad\qquad\qquad\qquad\qquad\qquad +\mC_{123}\Pi(-\q_2)\Pi(-\q_3)\Big].
\end{align}
Note that $[[\d q_2\d q_3]]$, as defined in \eqref{shorthand}, implicitly depends on $\q_1$ through the overall delta function expressing momentum conservation.

\subsection{Response functions}

Given a perturbative solution $\z$ of the classical equations of motion, we may formally expand $\Pi$ in terms of $\z$ to any given order in perturbation theory.  At quadratic order, we may thus write
\[
\Pi(\x_1)=\int\d^3\x_2\,\Omega(\x_2-\x_1)\z(\x_2)+\int\d^3\x_2\d^3\x_3\,\Lambda(\x_2-\x_1, \x_3-\x_1)\z(\x_2)\z(\x_3).
\]
where we will refer to the functions $\Omega$ and $\Lambda$ defined by this equation as {\it response functions}.
(Note we have made use here of the translation invariance of the background 3-geometry).
In momentum space, we then have
\[
\Pi(\q_1) = \Omega(-\q_1) \z(\q_1)+\int [[\d q_2 \d q_3]]\, \Lambda(\q_2, \q_3)\z(-\q_2)\z(-\q_3).
\]
Rotational invariance and momentum conservation (which in particular implies $2(\q_2\cdot\q_3)=q_1^2-q_2^2-q_3^2$)
imply that $\Omega$ and $\Lambda$ are scalar functions of the magnitudes $q_i$ such that
$\Omega(-\q_1)=\Omega(\q_1)=\Omega(q_1)$ and $\Lambda(\q_2,\q_3) = \Lambda(q_1, q_2, q_3)$.
Thus, in the following, we will simply write
\[
\label{response_fns}
\Pi(\q_1) = \Omega(q_1) \z(\q_1)+\int [[\d q_2 \d q_3]] \,\Lambda(q_i)\z(-\q_2)\z(-\q_3).
\]

Inserting this definition into \eqref{H_eom2} and expanding to quadratic order, making use of \eqref{H_eom1}, we find the response functions satisfy
\bea
\label{Omega_eom}
0 &=& \dot{\Omega}(q)+\frac{1}{2a^3\ep}\Omega^2(q)- 2\sigma a\ep q^2, \\
\label{Lambda_eom}
0 &=& \dot{\Lambda}(q_i) +\frac{1}{2a^3\ep}\big(\Omega(q_1)+\Omega(q_2)+\Omega(q_3)\big)
\Lambda(q_i) + \mathcal{X}(q_i),
\eea
where
\bea
\mathcal{X}(q_i) &=& 3\mA_{123}+\mB_{123}\Omega(q_1)+\mB_{213}\Omega(q_2)+\mB_{312}\Omega(q_3)+\mC_{123}\Omega(q_2)\Omega(q_3) \nn \\
&& +\mC_{213}\Omega(q_1)\Omega(q_3)+\mC_{312}\Omega(q_1)\Omega(q_2)+3\mD_{123}\Omega(q_1)\Omega(q_2)\Omega(q_3).
\eea

It is now straightforward to solve \eqref{Lambda_eom} perturbatively, starting from a solution of the linearised problem \eqref{Omega_eom}.
Specifically, given a solution $\z_q(z)$ of the linearised equation of motion
\[
\label{linear_eom}
0=\ddot{\z}_q+(3H+\dot{\ep}/\ep)\dot{\z}_q-\sigma a^{-2} q^2\z_q,
\]
it follows that 
\[
\Omega(q) = 2a^3\ep \dot{\z}_q/\z_q 
\]
is a solution of \eqref{Omega_eom}, and that
\[
\frac{\d}{\d z}\(\frac{1}{\z_q(z)}\) = -\frac{1}{2a^3\ep}\Omega(z,q)\(\frac{1}{\z_q(z)}\).
\]
The solution for $\Lambda$ is then
\[
\label{Lambda_soln}
\Lambda(z, q_i) = -\(\prod_i \frac{1}{\z_{q_i}(z)}\) \int^z_{z_0} \d z' \mathcal{X}(z',q_i)\prod_i\z_{q_i}(z'),
\]
where we will leave the lower limit $z_0$ in the integral unspecified for the time being.

\subsection{The domain-wall/cosmology correspondence}

Examining \eqref{bgd_eom}, \eqref{H_eom1} and \eqref{H_eom2} closely, we see that a perturbed {\it cosmological} solution expressed in terms of $\K^2$ and $\q_i$ analytically continues to a perturbed {\it domain-wall} solution expressed in terms of $\bK^2$ and $\vec{\bq}_i$, where
\[
\label{cont}
 \bK^2= -\K^2, \qquad \bq_i = -i q_i.
\]
The first continuation serves to reverse the sign of the potential in \eqref{bgd_eom} (taking, for example, dS to AdS), while the second ensures that $q_i^2=-\bq_i^2$, accounting for the necessary sign changes in the equations of motion \eqref{H_eom1} and \eqref{H_eom2} (specifically, in the coefficients $\mA$, $\mB$, $\mC$ and $\mD$, as well as in the first term on the r.h.s.~of \eqref{H_eom2}).
The choice of branch cut we made in this latter continuation ({\it i.e.}, $\bq_i=-iq_i$ rather than $\bq_i=+iq_i$) is determined by the necessity of mapping the cosmological Bunch-Davies vacuum behaviour, $\z \sim e^{-iq\tau}$ as $\tau \tto -\inf$ (where $\tau = \int \d z/a$), to the domain-wall solution that decays smoothly in the interior, $\z \sim e^{\bq\tau}$ as $\tau\tto-\inf$, as required for the computation of holographic correlation functions.

For the response functions, we see likewise that if we define $\Omega(q)$ and $\Lambda(q_i)$ to be the {\it cosmological} response functions with $\sigma=-1$, then the {\it domain-wall} response functions $\bOmega(\bq)$ and $\bLambda(\bq_i)$ are given by the simple analytic continuation
\[
\label{response_cont}
 \bOmega(\bq) = \bOmega(-iq) = \Omega(q), \qquad \bLambda(\bq_i)=\bLambda(-iq_i)=\Lambda(q_i).
\]
(Note that the response functions, as defined here, are independent of $\K^2$). 

In the remainder of this paper, we will use the unbarred variables $\K^2$, $q_i$ and the response functions $\Omega$ and $\Lambda$ when performing cosmological calculations, and the barred variables $\bK^2$, $\bq_i$ and response functions $\bOmega$ and $\bLambda$ for domain-wall calculations.  To analytically continue the results from domain-walls to cosmologies, and vice versa, we use \eqref{cont} and \eqref{response_cont}.

Finally, let us note the analytic continuations \eqref{cont} may equivalently be expressed in terms of QFT variables as
\[
\label{QFT_cont}
 \bN = -iN, \qquad \bq_i=-iq_i,
\]
where $\bN$ is the rank of the gauge group of the QFT dual to the domain-wall spacetime, and $N$ is the rank of the gauge group of the pseudo-QFT dual to the corresponding cosmology.  These relations follow from \eqref{cont}, noting that in the standard holographic dictionary $\bK^{-2} \propto \bN^2$, working in units where the AdS radius has been set to unity\footnote{\renewcommand{\baselinestretch}{1}\normalsize\small
In fact, in our later results we will see explicitly that holographic correlation functions calculated from the gravity side of the correspondence appear with an overall prefactor of $\bK^{-2}$.
On the QFT side of the correspondence, this prefactor corresponds to the overall prefactor of $\bN^2$ in correlators arising from the trace over gauge indices.}.
Our choice of branch cut in the continuation of $\bN$ ensures that the dimensionless effective QFT coupling, $g_{\mathrm{eff}}^2=g_{\mathrm{YM}}^2\bN/\bq = g_{\mathrm{YM}}^2 N/q$, does not change when we analytically continue from QFT to pseudo-QFT.  This is important because the QFT correlators may in general be non-analytic functions of $g_{\mathrm{eff}}^2$ at large $N$ \cite{Jackiw:1980kv, Appelquist:1981vg}.

\section{The cosmological bispectrum} 
\label{Bispec_section}

\subsection{Computation using response functions}

In this section we compute the 3-point function of cosmological curvature perturbations in terms of the second-order response function $\Lambda$.

We begin by quantising the interaction picture field $\z$ such that
\[
\hat{\z}(z,\x) = \int[\d q] \(\a(\q)\z_q(z)e^{i\q\cdot\x}+\a^\dag(\q)\z_q^*(z)e^{-i\q\cdot\x}\)
\]
(recalling that $z$ plays the role of proper time here), or equivalently, in momentum space,
\[
\hat{\z}(z,\q)=\a(\q)\z_q(z)+\a^\dag(-\q)\z_q^*(z).
\]
The creation and annihilation operators obey the usual commutation relations
\[
[\a(\q),\a^\dag(\q\,')] = (2\pi)^3 \delta(\q-\q\,').
\]
In these expressions, the mode function $\z_q(z)$ is a solution of the linearised equation of motion \eqref{linear_eom}, with initial conditions specified by the Bunch-Davies vacuum condition.

At tree level, the 3-point function in the in-in formalism may then be evaluated according to the standard formula \cite{Maldacena:2002vr}
\[
\label{zzz}
\<\hat{\z}(z,\q_1)\hat{\z}(z,\q_2)\hat{\z}(z,\q_3)\> = -i \int^z_{z_0} \d z' \< [{:}\hat{\z}(z,\q_1)\hat{\z}(z,\q_2)\hat{\z}(z,\q_3){:}\,,\,{:}\hat{H}^{(3)}(z'){:}]\>,
\]
where, to ensure convergence, a suitable infinitesimal rotation of the contour of integration is understood.  The lower limit $z_0$ represents some very early time (corresponding to $\tau$ very large and negative) at which the interactions are assumed to be switched on.  Note that both the operators appearing in the commutator in this formula are taken to be normal ordered as indicated.

Inserting the operator equivalent of \eqref{H3decomp} for $\hat{H}^{(3)}$ in the above formula, we may now proceed to evaluate the commutator explicitly, noting that for the cubic terms in $\hat{H}^{(3)}$ we may replace
\[
\hat{\Pi}(z,\q) = \a(\q)\Pi_q(z)+\a^\dag(-\q)\Pi^*_q(z) = \a(\q)\Omega(z,q)\z_q(z)+\a^\dag(-\q)\Omega^*(z,q)\z^*_q(z).
\]
In this manner, we find the full 3-point function
\begin{align}
\label{result1}
 \<\hat{\z}(z,\q_1)&\hat{\z}(z,\q_2)\hat{\z}(z,\q_3)\>  \nn \\[1ex]
&= - 4\K^{-2}(2\pi)^3\delta(\sum_i \q_i)\,\, \Im \Big[\Big(\prod_i \frac{1}{\z_{q_i}(z)}\Big)\int^z_{z_0} \d z' \mathcal{X}(z',q_i) \prod_i \z_{q_i}(z')\Big]\,\prod_i|\z_{q_i}(z)|^2 \nn \\[1ex]
&= (2\pi)^3\delta(\sum_i \q_i)\,\,4\K^{-2} \Im \left[\Lambda(z,q_i)\right]\,\prod_i |\z_{q_i}(z)|^2, 
\end{align}
where in the last line we have used \eqref{Lambda_soln}.  The lower limit of integration $z_0$ in \eqref{Lambda_soln} should thus be identified with the lower limit $z_0$ in \eqref{zzz}.  With this choice, we find $\Lambda \tto 0$ as $z\tto z_0$, consistent with the expected behaviour $\z(\q)\tto \z_q$ and $\Pi(\q)\tto \Omega(q)\z_q$ prior to the switching on of the interactions.

Introducing the notation
\begin{align}
\label{zeta_shorthand}
 \<\hat{\z}(z,\q_1)\hat{\z}(z,\q_2)\> & = (2\pi)^3\delta(\q_1+\q_2)\<\!\<\hat{\z}(z,q_1)\hat{\z}(z,-q_1)\>\!\>, \nn \\[1ex]
\<\hat{\z}(z,\q_1)\hat{\z}(z,\q_2)\hat{\z}(z,\q_3)\> &= (2\pi)^3\delta(\sum_i \q_i) \<\!\< \hat{\z}(z,q_1)\hat{\z}(z,q_2)\hat{\z}(z,q_3)\>\!\>,
\end{align}
the {\it bispectrum} of curvature perturbations $\<\!\< \hat{\z}(z,q_1)\hat{\z}(z,q_2)\hat{\z}(z,q_3)\>\!\>$
then satisfies
\[
\label{cosmo_result}
\frac{\<\!\<\hat{\z}(z,q_1)\hat{\z}(z,q_2)\hat{\z}(z,q_3)\>\!\>}{\prod_i \<\!\<\hat{\z}(z,q_i)\hat{\z}(z,-q_i)\>\!\>}= \Im \left[4\K^{-2}\Lambda(z,q_i)\right],
\]
where, as usual, the 2-point function is
\[
\<\!\<\hat{\z}(z,q_1)\hat{\z}(z,-q_1)\>\!\>=|\z_{q_1}(z)|^2.
\]
Noting that the linearised mode functions obey the Wronskian normalisation condition
\[
i\K^2 = \z_q\Pi_q^*-\Pi_q\z_q^* = -2i |\z_q(z)|^2\Im[\Omega(z,q)],
\]
we may express the 2-point function in terms of the response function $\Omega$,
\[ 
\label{cosmo_result2pt}
\<\!\<\hat{\z}(z,q_1)\hat{\z}(z,-q_1)\>\!\>=- \frac{\K^2}{2 \Im[\Omega(z,q)]}.
\]

Equations (\ref{cosmo_result2pt}) and (\ref{cosmo_result}) are the main result of this 
section: they express the power spectrum and the bispectrum in terms of response functions.
As we will see in the next section, these response functions (after analytic 
continuation) are directly related with 2- and 3-point functions of strongly coupled QFT via standard 
gauge/gravity duality.

\subsection{An example: slow-roll inflation}

As an illustration, let us use our results above to calculate the bispectrum to leading order in the slow-roll approximation.
As noted by Maldacena \cite{Maldacena:2002vr}, this calculation is most easily performed using the field redefinition
\[
\label{field_redef}
 \z = \z_c+ \big(\frac{\ddot{\vphi}}{2\dot{\vphi}H}+\frac{\ep}{4}\big)\z_c^2+\frac{\ep}{2}\D^{-2}(\z_c\D^2\z_c)+\ldots,
\]
where the dots indicate terms that vanish outside the horizon or are of higher order in slow roll.
The cubic action \eqref{Lagrangian} may then be rewritten to leading order in slow roll as 
\[
 S_c^{(3)} = \frac{1}{\K^2}\int \d^4 x\, 4\ep^2a^5 H \dot{\z}_c^2\D^{-2}\dot{\z}_c+\ldots
\]
Comparing with \eqref{Lagrangian}, we see that the field redefinition makes manifest the fact that the interaction is really of second order in the slow-roll parameter $\ep$.
The interaction Hamiltonian for $\z_c$ then has only the single term
\[
 \mD_c(q_i) = \frac{H}{6a^4\ep}\big(\frac{1}{q_1^2}+\frac{1}{q_2^2}+\frac{1}{q_3^2}\big).
\]

To evaluate the late-time value of the response function $\Lambda$, we use the formula \eqref{Lambda_soln}, substituting in the de Sitter solution
\[
 \z_q(\tau) \approx \frac{i\K H_*}{\sqrt{4\ep_*q^3}}(1+iq\tau)e^{-iq\tau}
\]
for the linearised mode functions.  Here, the asterisk indicates taking the value at the time of horizon crossing $z=z_*$ (where $q\approx a(z_*)H(z_*)$),
while the conformal time $\tau =\int\d z /a$.
We also use the fact that, to leading order in slow roll, the linear response function
\[
 \Omega_q(\tau) = \frac{2a^2\ep_*}{\z_q}\frac{\d\z_q}{\d\tau} = \frac{-2a\ep_*q^2}{H_*(1+iq\tau)},
\]
since $a\approx -1/H_*\tau$ and time derivatives of $\ep_*$ and $H_*$ are of higher order in slow roll.

We thus find
\[
\Lambda_0(q_i) = \frac{4\ep_*^2}{H_*^2}\big(\sum_{i>j}q_i^2q_j^2\big)\int^0_{-\inf}\d \tau' e^{-i(\sum_iq_i)\tau'}= i\frac{4\ep_*^2}{H_*^2}\frac{\sum_{i>j}q_i^2q_j^2}{\sum_i q_i},
\]
where the integration contour has been suitably rotated so as to ensure convergence of the lower limit and the subscript zero, both here and below,  denotes the value in the late-time limit $\tau\tto 0^-$.
Then,
\[
\<\!\<\hat{\z}_c(q_1)\hat{\z}_c(q_2)\hat{\z}_c(q_3)\>\!\>_0 =  \Im[4\K^{-2}\Lambda_0(\tau, q_i)]\Pi_i|\z_{0 q_i}(\tau)|^2
=\frac{H_*^4}{\K^2\ep_*(\prod_iq_i^3)}\frac{\sum_{i>j}q_i^2q_j^2}{4\sum_i q_i},
\]
in agreement with \cite{Maldacena:2002vr}, recalling that $\vphi$ here is dimensionless.
As in \cite{Maldacena:2002vr}, we then recover the 3-point function for $\z$ via the field redefinition \eqref{field_redef}.
This yields the usual result for approximately equal momenta, namely
\[
 \<\!\<\hat{\z}(q_1)\hat{\z}(q_2)\hat{\z}(q_3)\>\!\>_0 = \frac{H_*^4}{4\ep_*^2}\Big(\prod_i \frac{1}{2q_i^3}\Big)\Big[\frac{2\ddot{\vphi}_*}{\dot{\vphi}_*H_*}\sum_iq_i^3+
\ep_*\Big(\sum_iq_i^3+\sum_{i\neq j}q_iq_j^2+8\frac{\sum_{i>j}q_i^2q_j^2}{\sum_i q_i}\Big)\Big].
\]

\section{Holographic 3-point functions} 
\label{Hol_section}

\subsection{Holographic analysis}

Before commencing with our main holographic calculation in Section \ref{TTT_section}, let us first pause to briefly review some of the relevant background material for this calculation.

\subsubsection{Background solutions} 
\label{Bgds}

As mentioned briefly in the introduction to this paper, there are two general classes of domain-wall solutions for which a well understood holographic description exists.
We list these classes below: it is for these backgrounds that our holographic framework for cosmology is most readily applicable. 

\begin{itemize}
\item[(i)]{\it Asymptotically AdS domain-walls.}

In this case the solution behaves asymptotically as
\[
a(z) \sim e^{z}, \qquad \vphi \sim 0 \qquad {\rm as} \quad z \to \infty.
\]
The boundary theory has a UV fixed point which corresponds to the bulk AdS critical point.
Depending on the rate at which $\vphi$ approaches zero as $ z \to \infty$,
the QFT is either a deformation of the conformal field theory (CFT), or else the CFT in a state in which the dual scalar operator acquires a nonvanishing vacuum expectation value (see \cite{Skenderis:2002wp} for details).
Under the domain-wall/cosmology correspondence, these solutions are mapped to
cosmologies that are asymptotically de Sitter at late times.

\item[(ii)]{\it Asymptotically power-law solutions.}

In this case the solution behaves asymptotically as
\[
\label{power_law}
a(z) \sim (z/z_0)^n, \quad \vphi \sim \sqrt{2 n} \log (z/z_0) \quad {\rm as} \quad z \to \infty,
\]
where $z_0=n{-}1>0$.
In particular, for $n=7$ the asymptotic geometry is the near-horizon limit of a stack of D2 brane solutions.
In general, these solutions describe QFTs with a single dimensionful coupling
constant ({\it i.e.}, QFTs of the form \eqref{Lfree}), in the regime where the dimensionality of the coupling constant drives the dynamics \cite{Kanitscheider:2008kd}.
Under the domain-wall/cosmology correspondence, these solutions are mapped to cosmologies that
are asymptotically power-law at late times.

\end{itemize}

\subsubsection{Asymptotic analysis} 
\label{FG_section}

Holography relates bulk fields to local gauge-invariant operators of
the boundary QFT.  In particular, the bulk metric is related to the
boundary stress-energy tensor $T_{ij}$.  Bulk scalar fields, such as
the inflaton, correspond to boundary scalar operators ({\it e.g.},
$\mathrm{tr} F_{ij}F^{ij}$).
More precisely, the map is specified as follows.
First, recall that in order to define a quantum theory we must specify the behaviour of the fields
at infinity. In a gravitational theory, this means in particular
that the spacetime asymptotics must be prescribed. 
In gauge/gravity duality, the fields that specify the 
boundary conditions on the bulk side 
are identified with the sources of the boundary QFT operators 
\cite{Gubser:1998bc,Witten:1998qj}.
Correlation functions for these gauge-invariant operators may then be
extracted from the asymptotics of bulk solutions.  Conversely, given
the correlation functions of dual operators, one may reconstruct the
bulk asymptotics. 

Thus, to define the bulk theory, we need to specify appropriate boundary 
conditions. These boundary conditions must involve an arbitrary metric, 
since this will act as a source for the stress-energy tensor. Such 
boundary conditions are supplied by giving
an asymptotically locally AdS metric,
which in four dimensions takes the form\footnote{ \renewcommand{\baselinestretch}{1}\normalsize\small
In other spacetime dimensions, the general features of the
analysis remain the same although specific details differ.},
\begin{align}
\label{FG}
 \d s^2 &= \d r^2 + g_{ij}(r,x) \d x^i \d x^j, \nn\\
 g_{ij}(r,x) &= e^{2r}\big( g_{(0)ij}(x)+e^{-2r}g_{(2)ij}(x)+ \dots 
+ e^{-2 m r} g_{(2m)ij}(x) + \dots\big).
\end{align}
This encompasses the boundary conditions for the bulk metric, 
both for asymptotically AdS domain-walls and for asymptotically power-law 
solutions. In the former case,  
the radial coordinate $r$ may be identified with $z$, and
$2m=3$.  
For asymptotically power-law solutions,
one may perform a conformal transformation to the {\it dual frame}
\cite{Boonstra:1998mp} defined by $\tilde{g}_{ij} = \exp(-\lambda
\Phi) g_{ij}$, where $\lambda = \sqrt{2/n}$.  The asymptotic solution
above then describes the most general asymptotics for the dual frame
metric $\tilde{g}_{ij}$, where now $2m = (3n-1)/(n-1)>3$ and $r = \int
\exp(-\lambda\Phi/2) \d z$ (see \cite{Kanitscheider:2008kd} for
details).  In general, much of the holographic analysis for spacetimes
with power-law asymptotics may be obtained from that for
asymptotically $\mathrm{AdS}_{2m+1}$ spacetimes, which are related to
power-law spacetimes via a dimensional reduction on a
$\mathrm{T}^{2m-3}$ torus and analytic continuation in $m$
\cite{Kanitscheider:2009as}.

In the asymptotic expansion \eqref{FG}, the leading coefficient
$g_{(0)ij}(x)$ is 
an arbitrary (non-degenerate) three-dimensional
metric of the conformal boundary of the bulk spacetime.
Since this is
the metric on which the dual QFT lives, $g_{(0)ij}$ acts as the source
for the dual stress-energy tensor $T_{ij}$.  The subleading
coefficients $g_{(2k)ij}(x)$, with $k<m$, are then locally determined
in terms of $g_{(0)ij}$ via an asymptotic analysis of the field
equations.  The coefficient $g_{(2m)ij}(x)$, however, is only
partially constained by this asymptotic analysis. (On the QFT side,
these constraints correspond to the QFT Ward identities).  In fact, one
finds that the coefficient $g_{(2m)ij}(x)$ is directly related to the
expectation value of the boundary stress-energy tensor
\cite{deHaro:2000xn, Kanitscheider:2008kd}:
\[
\label{1pt_result}
 \<T_{ij}\>=\frac{1}{2\bar{\K}^2}(2m g_{(2m)ij}).
\]
An analogous relation also exists for the expectation value of the
dual scalar operator in terms of the asymptotic behaviour of the bulk
scalar field (see \cite{deHaro:2000xn, Kanitscheider:2008kd} for
details). We emphasize that this result only requires that Einstein
equations hold asymptotically.

Here, we focused our discussion on the stress-energy tensor.  
An analogous discussion holds for all operators: one should 
specify boundary conditions for the corresponding bulk fields
(and this is part of the definition of the theory).
If one includes such additional fields, then the holographic formulae 
such as (\ref{1pt_result}) will in general acquire additional 
terms \cite{BFS1,BFS2}, but the structure described above 
remains the same. More importantly for our purposes, 
since we are only interested in correlation functions of 
the stress-energy tensor, we only need to turn on a source for the stress-energy 
tensor, in which case the formulae above hold unchanged
(modulo contributions to (\ref{1pt_result}) from 
condensates of low-dimension operators, {\it cf.}~the discussion of the 
Coulomb branch flow in \cite{BFS1,BFS2}.
Such cases can be analysed along similar lines 
but we will not discuss this here).

The relation \eqref{1pt_result} may be read in two ways: (i) given a
bulk gravitational solution we may read off the dual QFT data encoded
by the solution; (ii) given QFT data we may reconstruct the bulk
asymptotic solution.  We stress that this asymptotic reconstruction is
possible even when gravity is strongly coupled in the interior.  
The coefficients up to
$g_{(2m)ij}$ just encode the boundary conditions, {\it i.e.}, the fact that
we are considering asymptotically locally AdS configurations (in the
dual frame for the power-law case).
In gauge/gravity duality, these
terms encode the fact that we have turned on a source for the dual
operator (the stress-energy tensor for the case at hand) and this is
unrelated to whether the dual QFT is at weak or strong coupling.
The first term to depend on the bulk dynamics is $g_{(2m)ij}$. When
gravity is weakly coupled, this coefficient is determined by the
behaviour of the gravitational solution deep in the interior.  When
gravity is strongly coupled, this coefficient should be obtained by
solving the full stringy dynamics in the interior.  Gauge/gravity
duality requires that the value obtained this way {\it must agree}
with the $g_{(2m)ij}$ determined via \eqref{1pt_result} from the
weakly coupled dual QFT.

\subsubsection{Radial Hamiltonian formulation}

In the following, rather than using \eqref{1pt_result} directly, we
will instead employ the radial Hamiltonian formulation of
\cite{Papadimitriou:2004ap, Papadimitriou:2004rz}.  Here, the radial
direction plays a role equivalent to that of time in the usual
Hamiltonian formalism.  The radial Hamiltonian formulation has a
number of advantages for our present purposes; in particular, 
it leads to a universal formula for the 1-point function that 
is independent of any of the issues (additional fields, {\it etc.}) 
discussed in the previous subsection. It also
permits us to work with an arbitrary potential for the scalar field,
so long as this potential admits background solutions of either the
asymptotically AdS or asymptotically power-law form\footnote{\renewcommand{\baselinestretch}{1}\normalsize\small
In contrast, the formula \eqref{1pt_result} must be established on a
  case by case basis for different potentials, as in
  \cite{deHaro:2000xn,BFS1,BFS2}.}.

A key feature of spacetimes of the form \eqref{FG} is that, to leading order as $r\tto \inf$,  the radial derivative is equal to the dilatation operator $\delta_D$, {\it i.e.},
\[
\label{RGreln}
 \partial_r = \delta_D(1+O(e^{-2r})), \qquad
 \delta_D = \int \d^3x \Big( 2g_{ij}\frac{\delta}{\delta g_{ij}}+(\Delta-3)\Phi\frac{\delta}{\delta\Phi}\Big),
\]
taking the bulk field $\Phi$ to be dual to a scalar operator of weight $\Delta$.
This equivalence allows one to trade the asymptotic radial expansion \eqref{FG} for a covariant expansion in eigenfunctions of the dilatation operator.  By definition, an eigenfunction $A_{(n)}$ of weight $n$ satisfies
\[
 \delta_D A_{(n)} = -n A_{(n)}.
\]
From \eqref{RGreln}, $A_{(n)} \sim e^{-nr}(1+O(e^{-2r}))$, so the radial expansion and the expansion in eigenfunctions of the dilatation operator are closely related.  The latter expansion is manifestly covariant, however, whereas expanding in the bulk radial coordinate is not a covariant operation.

In the radial Hamiltonian formalism then, the expectation value of the
dual stress-energy tensor is given by
\[
\label{Radial_formula}
 \<T^i_j\> = \(\frac{-2}{\sqrt{g}}{\Pi^i_j}\)_{(3)}
\]
where $\Pi_{ij}$ is the radial canonical momentum in `synchronous'
(Fefferman-Graham)
gauge where $N_i=0$ and $N=1$, and the subscript indicates taking the piece
with overall dilatation weight\footnote{\renewcommand{\baselinestretch}{1}\normalsize\small
In odd bulk dimensions the
  transformation of this specific coefficient also has an additional
  anomalous contribution due to the conformal anomaly
  \cite{Henningson:1998gx}.  In our case there is no
  anomaly, however, and this coefficient is a true eigenfunction of
  $\delta_D$.} three. This is the universal formula we alluded to above\footnote{\renewcommand{\baselinestretch}{1}\normalsize\small
While (\ref{Radial_formula}) holds universally, expressing $\Pi^i_j$ 
in terms of the coefficients in the asymptotic expansion of the bulk 
fields depends on the details of theory under consideration (field
content, interactions, {\it etc.}). Fortunately, we will not need this information 
here.}. 
To extract the piece with dilatation weight three,
$\Pi_{ij}$ may first be decomposed in
eigenfunctions of the dilatation operator.  In general, the radial
canonical momentum will contain pieces with weight less than three: the
process of holographic renormalisation then amounts to determining
these terms through the asymptotic analysis and subtracting
them. 
In \cite{Papadimitriou:2004rz, Papadimitriou:2004ap}, it is shown that
removing these pieces is equivalent to adding local boundary covariant
counterterms to the on-shell action.

For asymptotically AdS domain-walls, the radial canonical momentum is
\[
 \Pi^i_j = \frac{1}{2\bK^2}\sqrt{g}(K^i_j-K\delta^i_j),
\]
where $K_{ij}=(1/2)\partial_z g_{ij}$ is the extrinsic curvature of constant-$z$ slices. (Recall for domain-walls, the $z$ coordinate is a radial variable).
In the case of asymptotically power-law domain-walls, the relevant radial canonical momentum is instead that of the dual frame \cite{Kanitscheider:2008kd}, namely
\[
\tilde{\Pi}^i_j = \frac{1}{2\bK^2}\sqrt{\tilde{g}}e^{\lambda\Phi}(\tilde{K}^i_j-(\tilde{K}+\lambda\Phi_{,r})\delta^i_j).
\]
Here, all tilded quantities belong to the dual frame and $\D_r = e^{\lambda\vphi/2}\D_z$.  (Note the r.h.s.~of \eqref{Radial_formula} should also be evaluated in the dual frame).

\subsubsection{3-point functions}

Starting with the 1-point function in the presence of sources, $\<T_{ij}\>_s$ (given by \eqref{Radial_formula} above), higher correlation functions may be obtained through repeated functional differentiation with respect to the source $g_{(0)kl}$, after which the source is set to its background value. 
In performing this operation, one must be careful to note that the stress-energy tensor $T_{ij}$ has itself a purely classical dependence on the metric; this additional metric dependence gives rise to contact terms, some of which we need to keep track of.  Specifically, when computing the 3-point function, we need to retain {\it semi-local} contact terms in which only two of the three points involved are coincident, since terms of this form contribute to local-type non-Gaussianity as we will see later.  We may, on the other hand, discard {\it ultralocal} contact terms in which all three points are coincident: such terms are generically scheme-dependent, {\it i.e.}, they may be removed by the addition of finite local counterterms.
The same is not true for the semi-local terms, which cannot be altered by finite local counterterms.

Expanding the 1-point function in the presence of sources to quadratic order about a flat background, we have
\begin{align}
\delta\<T_{ij}(\x_1)\>_s &= \int \d^3\x_2 \frac{\delta\<T_{ij}(\x_1)\>}{\delta g^{kl}(\x_2)}\Big|_{0}\delta g^{kl}(\x_2)+\half\int \d^3\x_2\d^3\x_3
\frac{\delta^2\<T_{ij}(\x_1)\>}{\delta g^{kl}(\x_2)\delta g^{mn}(\x_3)}\Big|_{0}\delta g^{kl}(\x_2)\delta g^{mn}(\x_3) \nn \\[2ex]
&= -\half\int\d^3\x_2\<T_{ij}(\x_1)T_{kl}(\x_2)\>\delta g^{kl}(\x_2) \nn \\[1ex]
&\quad +\frac{1}{8}\int\d^3\x_2\d^3\x_3 \Big[ \<T_{ij}(\x_1)T_{kl}(\x_2)T_{mn}(\x_3)\>
+\delta(\x_2-\x_3)\<T_{ij}(\x_1)T_{kl}(\x_2)\>\delta_{mn} \nn \\[1ex]
& \qquad \qquad  -2\<T_{ij}(\x_1)\frac{\delta T_{kl}(\x_2)}{\delta g^{mn}(\x_3)}\>-4\<\frac{\delta T_{ij}(\x_1)}{\delta g^{mn}(\x_3)}T_{kl}(\x_2)\>
\Big]\delta g^{kl}(\x_2)\delta g^{mn}(\x_3), 
\end{align}
where the zero subscripts in the first line indicate setting the sources to their background value ({\it i.e.}, setting $g_{ij}=\delta_{ij}$),
and in the second line we have dropped all ultralocal contact terms, but retained those where only two points are coincident.

Setting $g_{ij}=(1-2\psi)\delta_{ij}$ (so that $g^{ij}=\delta^{ij}+\delta g^{ij} = (1+2\psi+4\psi^2)\delta^{ij}$),
and noting that $\<T_{ij}(\x_1)\>_0=0$,
the expansion of $\<T^i_i(\x_1)\>_s$ to quadratic order in the source $\psi$ is
\begin{align}
\label{psi_exp}
\delta\<T^i_i(\x_1)\>_s &=
(\delta^{ij}+\delta g^{ij}(\x_1))\delta\<T_{ij}(\x_1)\>_s \nn \\[1ex]
&= -\int\d^3\x_2\<T(\x_1)T(\x_2)\>\psi(\x_2) \\[1ex]
&\quad +\int\d^3\x_2\d^3\x_3 \Big[\half \<T(\x_1)T(\x_2)T(\x_3)\>  
 -\half\<T(\x_1)T(\x_2)\>\delta(\x_2-\x_3) \nn\\[1ex]
& \quad -2\<T(\x_1)T(\x_2)\>\delta(\x_1-\x_3) -\<T(\x_1)\Upsilon(\x_2,\x_3)\> -2\<T(\x_2)\Upsilon(\x_1,\x_3)\>
\Big]\psi(\x_2)\psi(\x_3), \nn
\end{align}
where the operators
\[
\label{Upsilon_def}
T(\x) = \delta^{ij}T_{ij}(\x), \qquad \Upsilon(\x_1,\x_2) = \delta^{ij}\delta^{kl}\frac{\delta T_{ij}(\x_1)}{\delta g^{kl}(\x_2)}\Big|_{0} .
\]
Note that $\Upsilon$ is symmetric under interchange of its arguments, $\Upsilon(\x_1,\x_2)=\Upsilon(\x_2,\x_1)$, since the definition above is equivalent to
\[
 \Upsilon(\x_1,\x_2) =2\delta^{ij}\delta^{kl}\frac{\delta^2 S}{\delta g^{ij}(\x_1)\delta g^{kl}(\x_2)}\Big|_{0}+\frac{3}{2}T(\x_1)\delta(\x_1-\x_2).
\]

In momentum space, we then have
\begin{align}
\label{psi_exp2}
\delta\<T^i_i(\vbq_1)\>_s =
&-\<\!\<T(\bq_1)T(-\bq_1)\>\!\> \psi(\vbq_1) \\[1ex]
&+\int[[\d\bq_2 \d\bq_3]] \Big[\half \<\!\<T(\bq_1)T(\bq_2)T(\bq_3)\>\!\> -\half\<\!\<T(\bq_1)T(-\bq_1)\>\!\> -2\<\!\<T(\bq_2)T(-\bq_2)\>\!\> \nn \\[1ex]
&\qquad\qquad\qquad\qquad   -\<\!\<T(\bq_1)\Upsilon(\bq_2,\bq_3)\>\!\> -2\<\!\<T(\bq_2)\Upsilon(\bq_1,\bq_3)\>\!\>
\Big]\psi(-\vbq_2)\psi(-\vbq_3), \nn
\end{align}
where $[[\d\bq_2 \d\bq_3]]$ is defined analogously to in \eqref{shorthand}, and we have introduced the shorthand
\bea
\label{T_shorthand}
\<T(\vbq_1)T(\vbq_2)\> &=& (2\pi)^3\delta(\vbq_1+\vbq_2)\<\!\<T(\bq_1)T(-\bq_1)\>\!\>, \nn \\[1ex]
\<T(\vbq_1)\Upsilon(\vbq_2,\vbq_3)\> &=& (2\pi)^3\delta(\sum_i\vbq_i)\<\!\<T(\bq_1)\Upsilon(\bq_2,\bq_3)\>\!\>, \nn \\[1ex]
 \<T(\vbq_1)T(\vbq_2)T(\vbq_3)\> &=& (2\pi)^3\delta(\sum_i\vbq_i) \<\!\<T(\bq_1)T(\bq_2)T(\bq_3)\>\!\>.
\eea

\subsection{Computation of \texorpdfstring{$\<TTT\>$}{<TTT>}}
\label{TTT_section}

We are now ready to compute the 3-point function for the trace of the stress-energy tensor of the holographically dual QFT, considering first asymptotically AdS and then asymptotically power-law domain-wall backgrounds.  Ultimately, we will see how the result may be simply expressed in terms of the domain-wall response functions $\bOmega$ and $\bLambda$.

Our basic strategy will be to expand the dual 1-point function in the presence of sources to quadratic order in $\psi$, as in \eqref{psi_exp}.  The raw ingredients for the calculation will be the Hamiltonian and momentum constraint equations (given in Appendix \ref{App_constraints}); the equation of motion \eqref{H_eom2}; the gauge-invariant definition of $\z$, \eqref{zeta_gi}; and the definition \eqref{response_fns} of the response functions.

\subsubsection{Asymptotically AdS case}

Working in synchronous (Fefferman-Graham) gauge where $N_i=0$ and $N=1$, for asymptotically AdS domain-walls we have
\[
\label{1ptfn}
 \delta\<T^i_i(x)\>_s = \frac{2}{\bK^2}\delta K_{(3)} = \frac{1}{\bK^2}(\dot{h}-h_{ij}\dot{h}_{ij})_{(3)}
\]
where $h=h_{ii}$.
We now wish to expand $\delta\<T^i_i(x)\>_s$ to quadratic order in $\psi$.

Let us start by reviewing the computation of $\delta\<T^i_i(x)\>_s$ to linear order in $\psi$.
To this end, we note first that, at linear order, the Hamiltonian and momentum constraints read\footnote{\renewcommand{\baselinestretch}{1}\normalsize\small
The full constraint equations expanded to second order may be found in Appendix \ref{App_constraints}.  Since we are in synchronous gauge here, we set $\dN=0$.}
\[ \label{const_linear}
  \dot{\psi} = (\ldots)\delta\vphi, \qquad \dot{h} = -\frac{2\bq^2}{a^2H}\psi+\frac{\dot{\vphi}}{H}\delta\dot{\vphi}+(\ldots)\delta\vphi,
\qquad \dot{\omega}_i = 0.
\]
Setting  the tensors $\g_{ij}$ to zero in (\ref{metric_perts}) (noting they decouple at linear order), 
we obtain
 \[
 \dot{h}_{ij}=\frac{\bq_i\bq_j}{\bq^2}\dot{h} + (\ldots)\delta\vphi.
 \]
It follows that, in order to obtain $\delta\<T^i_i(x)\>_s$ to linear order in $\psi$, we need to find the 
relation between $\delta\dot{\vphi}$ and $\psi$ to linear order. This is obtained by using the definition of 
the response function, along with \eqref{H_eom1} and the definition \eqref{zeta_gi} of $\z$ to linear order.
On one hand, we have
\[ \label{zdot_response}
 \dot{\z}=\frac{1}{2a^3\ep}\Pi=\frac{1}{2a^3\ep}\bOmega(\bq)\z = -\frac{1}{2a^3\ep}\bOmega(\bq)\psi + (\ldots)\delta\vphi,
 \]
and on the other hand,
\[ \label{zdot_linear}
 \dot{\z} = (-\psi-\frac{H}{\dot{\vphi}}\delta\vphi)\dot{} =-\frac{H}{\dot{\vphi}}\delta\dot{\vphi}+(\ldots)\delta\vphi.
 \]
Thus, at linear order,
\[
\label{rules}
 \delta\dot{\vphi} = \frac{H}{a^3\dot{\vphi}}\bOmega(\bq)\psi+(\ldots)\delta\vphi, \qquad \dot{h}_{ij}= 
\frac{\bq_i\bq_j}{\bq^2}\(\frac{\bOmega(\bq)}{a^3}-\frac{2\bq^2}{a^2H}\)\psi+(\ldots)\delta\vphi.
\]
This is all that we need in order to derive the 2-point function (as we will do below). Moreover,
in the calculations to follow, we will use these results to replace all $\delta\dot{\vphi}$ and $\dot{h}_{ij}$ terms appearing in quadratic combinations.

We will now do the computation to quadratic order. The steps involved are the same; the main difference 
is that there are quadratic sources that are processed using (\ref{rules}). 
Let us start by examining the full Hamiltonian constraint at quadratic order.  From \eqref{Ham_constr},
in position space we have
\[
 (\dot{h}-h_{ij}\dot{h}_{ij}) = \frac{1}{2H}(R_{(1)}+R_{(2)})+\frac{\dot{\vphi}}{H}\delta\dot{\vphi}
-\frac{1}{8H}\dot{h}^2+\frac{1}{8H}\dot{h}_{ij}\dot{h}_{ij}+\frac{1}{2H}\delta\dot{\vphi}^2 +(\ldots)\delta\vphi+ (\ldots)\delta\vphi^2.
\]
The spatial curvature terms $R_{(1)}$ and $R_{(2)}$ are simply local functions of $\psi$, however, (for example,
$R_{(1)}=4a^{-2}\D^2\psi$) and so holographically these terms contribute only ultralocal contact terms to $\delta\<T^i_i(x)\>_s$.
We may therefore discard these terms immediately.  The remaining quadratic terms may then be replaced using \eqref{rules}.  Up to ultralocal contact terms, in momentum space this gives
\begin{align}
\label{eq1}
(\dot{h}-h_{ij}\dot{h}_{ij})(\vbq_1)  =  \frac{\dot{\vphi}}{H}\delta\dot{\vphi}(\vbq_1)+
\int & [[\d\bq_2\d\bq_3]]
 \Big[\frac{1}{8a^6H}\Big(\frac{2}{\ep}-1+\frac{(\vbq_2\cdot\vbq_3)^2}{\bq_2^2\bq_3^2}\Big)\bOmega(\bq_2)\bOmega(\bq_3) \nn \\[1ex]
& +\frac{1}{2a^5H^2}\Big(\bq_3^2-\frac{(\vbq_2\cdot\vbq_3)^2}{\bq_2^2}\Big)\bOmega(\bq_2)
\Big]\psi(-\vbq_2)\psi(-\vbq_3) + \ldots
\end{align}
where $[[\d\bq_2\d\bq_3]]$ is defined as in \eqref{shorthand}.

We now need to express $\delta\dot{\vphi}$ in terms of $\psi$, working to quadratic order.
Firstly, from the gauge-invariant definition \eqref{zeta_gi} of $\z$, in synchronous gauge we have
\begin{align}
\z  &= -\psi -\psi^2 + \ldots,  \nn\\[1ex]
\dot{\z} &= -\dot{\psi}-\frac{H}{\dot{\vphi}}\delta\dot{\vphi}-2\psi\dot{\psi}
+\frac{H}{\dot{\vphi}^2}\delta\dot{\vphi}\delta\dot{\vphi} \nn \\[1ex]
&\quad +\frac{1}{4}\pi_{ij}\Big[
-\frac{\delta\dot{\vphi}}{\dot{\vphi}}\dot{h}_{ij}
-2(\dot{\chi}_{,ki}+\dot{\omega}_{k,i})(-2\psi\delta_{jk})
-(\dot{\chi}_{,k}+\dot{\omega}_k)(-2\psi_{,k}\delta_{ij})\Big]  + \ldots,
\end{align}
where we have omitted terms that vanish when the sources are restricted to $h_{ij}=-2\psi\delta_{ij}$, $\delta\vphi=0$.
Upon replacing time-derivatives of perturbations in the quadratic terms using \eqref{rules}, we then find
\begin{align}
\label{eq2}
\dot{\z}(\vbq_1) =  -\dot{\psi}(\vbq_1)-\frac{H}{\dot{\vphi}}\delta\dot{\vphi}(\vbq_1)+& \int  [[\d\bq_2\d\bq_3]] \Big[
\frac{1}{8a^6\ep H}\Big(\frac{2}{\ep}-1+\frac{(\vbq_1\cdot\vbq_3)^2}{\bq_1^2\bq_3^2}\Big)\bOmega(\bq_2)\bOmega(\bq_3) \nn \\[1ex]
& +\frac{1}{2a^3}\Big(1+\frac{(\vbq_2\cdot\vbq_3)}{\bq_2^2}-\frac{(\vbq_1\cdot\vbq_2)^2}{\bq_1^2\bq_2^2}
+\frac{1}{a^2\dot{\vphi}^2}\Big(\bq_3^2-\frac{(\vbq_1\cdot\vbq_3)^2}{\bq_1^2}\Big)\Big)\bOmega(\bq_2) \nn \\[1ex]
&  +\frac{1}{a^2H}\frac{(\vbq_1\cdot\vbq_2)^2}{\bq_1^2}
\Big]\psi(-\vbq_2)\psi(-\vbq_3) + \ldots,
\end{align}
again dropping ultralocal contact terms. This is the analogue of (\ref{zdot_linear}) at quadratic order.

At linear order, the momentum constraint implied that $\dot{\psi}$ is proportional to $\delta \varphi$
(first equation in (\ref{const_linear})). To quadratic order we get
\begin{align}
\dot{\psi} &= -\frac{1}{4}(-2\psi\delta_{ij})\dot{h}_{ij}
+\D^{-2}\D_i\Big[
\frac{1}{4}\D_{j}\big((-2\psi\delta_{jk})\dot{h}_{ki}\big) \nn\\ &\qquad\qquad\qquad\qquad\qquad\qquad
+\frac{1}{8}\dot{h}_{jk}(-2\psi_{,i}\delta_{jk})-\frac{1}{8}\dot{h}_{ij}(-6\psi_{,j})\Big]
+\ldots, 
\end{align}
where again we omit terms that vanish when the sources are set to $h_{ij}=-2\psi\delta_{ij}$, $\delta\vphi=0$.
Using \eqref{rules} for the quadratic terms, we then find
\begin{align}
\label{eq3}
\dot{\psi}(\vbq_1) = \int[[\d\bq_2\d\bq_3]] &\Big[\frac{1}{2a^3}\Big(1+\frac{(\vbq_1\cdot\vbq_3)}{2\bq_1^2}-\frac{(\vbq_1\cdot\vbq_2)^2}{\bq_1^2\bq_2^2}
-\frac{3(\vbq_1\cdot\vbq_2)(\vbq_2\cdot\vbq_3)}{2\bq_1^2\bq_2^2}\Big)\bOmega(\bq_2) \\[1ex]
&+\frac{1}{a^2H}\Big(\frac{(\vbq_1\cdot\vbq_2)^2}{\bq_1^2}-\frac{(\vbq_1\cdot\vbq_3)\bq_2^2}{2\bq_1^2}+\frac{3(\vbq_1\cdot\vbq_2)(\vbq_2\cdot\vbq_3)}{2\bq_1^2}\Big)
\Big]\psi(-\vbq_2)\psi(-\vbq_3) + \ldots \nn
\end{align}
We now work out the analogue of (\ref{zdot_response}) to quadratic order.
From \eqref{H_eom1} we obtain,
\begin{align}
\label{eq4}
\dot{\z}(\vbq_1) = -\frac{1}{2a^3\ep}\bOmega(\bq_1)\psi(\vbq_1) +\int[[\d\bq_2\d\bq_3]] \Big[&
\frac{1}{2a^3\ep}(\bLambda(\bq_i)-\bOmega(\bq_1)) +\mB_{123} +2\mC_{312}\bOmega(\bq_2) \nn \\ & +3\mD_{123}\bOmega(\bq_2)\bOmega(\bq_3)\Big]
\psi(-\vbq_2)\psi(-\vbq_3) + \ldots,
\end{align}
where we need retain only the semilocal contact terms appearing in $\mB_{123}$.

Finally, we may now combine \eqref{eq1}, \eqref{eq2}, \eqref{eq3} and \eqref{eq4}, symmetrise under $\vbq_2\leftrightarrow\vbq_3$, and substitute for $\mB_{123}$, $\mC_{312}$ and $\mD_{123}$ using \eqref{Cdef} and \eqref{Ddef}.  After many cancellations, we are left with the simple result
\begin{align}
\label{hdot_result}
 (\dot{h}-h_{ij}\dot{h}_{ij})(\vbq_1) = \frac{1}{a^3}\bOmega(\bq_1)\psi(\vbq_1)  
+ \int &[[\d\bq_2 \d\bq_3]] \frac{1}{a^3}\big[ {-}\bLambda(\bq_i) +\bOmega(\bq_1) \nn\\[1ex]&\qquad\quad
+\frac{3}{2}\big(\bOmega(\bq_2)+\bOmega(\bq_3)\big)\big]\psi({-}\vbq_2)\psi({-}\vbq_3) + \ldots
\end{align}
According to \eqref{1ptfn}, we must now extract the piece with dilatation weight three.
In the present case, with $g_{ij}$ and $\Phi$ as given in \eqref{gij_def} and \eqref{fullphi_def}, 
the dilatation operator \eqref{RGreln} evaluates to
\[
 \delta_D = a\frac{\partial}{\partial a} + (\Delta-3)\Big(\vphi\frac{\partial}{\partial \vphi} +\int\d^3x \delta\vphi\frac{\delta}{\delta \delta\vphi}\Big). 
\]
As the terms of interest in \eqref{hdot_result} are independent of $\delta\vphi$, however, 
their eigenfunction expansion under the full dilatation operator simply coincides with that obtained using the background dilatation operator.
Noting that the scale factor $a$ has dilatation weight minus one, we then obtain
\begin{align}
 \delta\<T^i_i(\vbq_1)\>_s  &= \bK^{-2} \bOmega_{(0)}(\bq_1)\psi_{(0)}(\vbq_1) 
+\int [[\d\bq_2 \d\bq_3]] \bK^{-2}\big[
{-}\bLambda_{(0)}(\bq_i)+\bOmega_{(0)}(\bq_1) \nn\\[1ex] &\qquad\qquad\qquad\qquad\qquad
 {+}\frac{3}{2}\big(\bOmega_{(0)}(\bq_2)+\bOmega_{(0)}(\bq_3)\big)\big]\psi_{(0)}({-}\vbq_2)\psi_{(0)}({-}\vbq_3) + \ldots 
\end{align}
where the dots indicate terms that depend other sources. 
Comparing with \eqref{psi_exp2}, we then see that
\begin{align}
\label{holo_result1}
-\bK^{-2}\bOmega_{(0)}(\bq) &= \<\!\<T(\bq)T(-\bq)\>\!\>, \\[1ex]
\label{holo_result2}
-\bK^{-2}\bLambda_{(0)}(\bq_i) &= \half \<\!\<T(\bq_1)T(\bq_2)T(\bq_3)\>\!\>+\sum_i \half\<\!\<T(\bq_i)T(-\bq_i)\>\!\> \nn\\
& \qquad -\big[\<\!\<T(\bq_1)\Upsilon(\bq_2,\bq_3))\>\!\>+\<\!\<T(\bq_2)\Upsilon(\bq_1,\bq_3))\>\!\>+\<\!\<T(\bq_3)\Upsilon(\bq_1,\bq_2))\>\!\>\big].
\end{align}
This is the main result of the present section: we have obtained holographic formulae for 2- and 3-point functions
of the trace of the dual stress-energy tensor along a general holographic RG flow 
in terms of the response functions $\bOmega$ and $\bLambda$. While the formula for the 2-point function
is of course already known \cite{Papadimitriou:2004rz}, the result for the 3-point function is new and should 
have applications beyond the current work. Note that the terms in (\ref{holo_result2}) involving 2-point functions
do not contribute when all three operators are at separated points. These formulae were derived above 
for asymptotically AdS domain-walls but we shall see in the next subsection that they also hold in the case of 
asymptotically power-law domain-walls.

In these formulae, the subscript zero indicates taking the piece of the response functions with zero weight under dilatations.
To extract this piece correctly, one first expands the response functions in eigenfunctions of the dilatation operator, then determines the terms with eigenvalues less than zero through an asymptotic analysis of the response function equations of motion \eqref{Omega_eom} and \eqref{Lambda_eom}.  (In this asymptotic analysis, one replaces radial derivatives with the dilatation operator according to \eqref{RGreln}, and then collects together terms of equal dilatation weight).  The weight zero pieces of the response functions are then obtained by subtracting these terms with negative dilatation weight from the full response functions and taking the limit $z\tto\inf$.  (For an explicit worked example,
see Section 4.3 of \cite{McFadden:2010na}).  The relevant issue here is that the subtraction of terms with negative weight (which diverge as $z\tto \inf$) may induce a change in the zero weight (finite) part as well.

Fortunately, however, we are saved from having to carry out any of this analysis in detail
by virtue of the fact that the cosmological formula \eqref{cosmo_result} for the bispectrum involves taking the imaginary part of the cosmological response function at late times.  The counterterms one subtracts to obtain the weight-zero piece of the domain-wall response functions through the procedure described above are all analytic functions of $\bq^2$, and hence under the continuation $\bq^2=-q^2$, these terms remain real and so do not contribute to the imaginary part of the cosmological response functions.

\subsubsection{Asymptotically power-law case}

The holographic calculation for the case of asymptotically power-law domain-walls is very closely related to the calculations above for asymptotically AdS domain-walls.

The perturbed dual frame metric, when written in synchronous gauge so that the dual frame lapse and shift perturbations vanish, reads
\[
 \d \tilde{s}^2 = e^{-\lambda\Phi}\d s^2 = \d r^2+\tilde{a}^2(\delta_{ij}+\tilde{h}_{ij})\d x^i \d x^j,
\]
where $\tilde{a} = ae^{-\lambda\vphi/2}$ and $\d r = e^{-\lambda\vphi/2}\d z$.
The dual frame metric perturbations $\tilde{h}_{ij}=-2\tilde{\psi}\delta_{ij}+2\tilde{\chi}_{,ij}+2\tilde{\omega}_{(i,j)}+\tilde{\g}_{ij}$ may be expressed in terms of their Einstein frame counterparts through the relations
\begin{align}
\label{DF2EF}
&\tilde{\psi} = \psi+\frac{\lambda}{2}\delta\vphi -\lambda\delta\vphi\psi-\frac{\lambda^2}{4}\delta\vphi^2, \qquad
&\tilde{\omega}_i = (1-\lambda\delta\vphi)\omega_i, \nn \\
&\tilde{\chi} = (1-\lambda\delta\vphi)\chi, \qquad
&\tilde{\g}_{ij} = (1-\lambda\delta\vphi)\g_{ij}.
\end{align}
Note that while the Einstein frame shift vanishes ($\dN_i=0$), there is a nonzero Einstein frame lapse perturbation
\[
\label{nusub}
\dN = (\lambda/2)\delta\vphi+(\lambda^2/8)\delta\vphi^2.
\]

The 1-point function in the presence of sources is given by the canonical momentum in the dual frame,
\[
\label{DF1ptfn}
\<T^i_i(x)\>_s = \Big[\bK^{-2}e^{\lambda\Phi}(2\tilde{K}+3\lambda\Phi_{,r})\Big]_{(3)},
\]
where $\tilde{K}_{ij}=(1/2)\D_r\tilde{g}_{ij}$.
As in the case of asymptotically AdS domain-walls, we wish to expand the 1-point function in the presence of sources to quadratic order in  $\tilde{\psi}$.
From \eqref{DF2EF}, however, expanding in powers of $\tilde{\psi}$ in the dual frame is equivalent to expanding in powers of $\psi$ in the Einstein frame
({\it i.e.}, the coefficients of $\tilde\psi$ and $\tilde\psi^2$ in the dual frame equal the coefficients of $\psi$ and $\psi^2$ in the Einstein frame).
Expanding \eqref{DF1ptfn} and converting dual frame perturbations into their Einstein frame equivalents, therefore, we find
\begin{align}
\delta \big[\bK^{-2}e^{\lambda\Phi}(2\tilde{K}+3\lambda\Phi_{,r})\big] =
\bK^{-2}e^{3\lambda\vphi/2}\big(& \dot{h}-h_{ij}\dot{h}_{ij}+\ldots\big),
\end{align}
where we have omitted all terms that do not contribute to the expansion in $\psi$.

In principle, one would now proceed as in the previous section, by expanding out the Hamiltonian and momentum constraints in the Einstein frame, using response functions to substitute for the radial derivatives of metric perturbations where necessary: the only difference here being that there is now a nonzero lapse perturbation $\dN$.
Fortunately, however, there is no need to repeat these calculations, since upon inspection of the constraint equations \eqref{Ham_constr} and \eqref{mom_constr}, we see that only the lapse perturbation $\dN$ and its spatial derivatives appear, and never the radial derivative $\delta\dot{N}$.
Thus, from \eqref{nusub}, these additional terms involving the lapse perturbation do not contribute to the expansion of the 1-point function in powers of $\psi$.  Similarly, the gauge-invariant variable $\z$ does not involve $\dN$, as may be seen from \eqref{zeta_gi}.
We may therefore straightforwardly lift the result \eqref{hdot_result} from the previous section, whence
\begin{align}
& \delta \big[\bK^{-2}e^{\lambda\Phi}(2\tilde{K}+3\lambda\Phi_{,r})\big](\vbq_1)  \\[1ex]
& = \frac{1}{\bK^2 \tilde{a}^3}\bOmega(\bq_1)\psi(\vbq_1) 
  + \int[[\d\bq_2\d\bq_3]]\, \frac{1}{\bK^2 \tilde{a}^3}\big[
{-}\bLambda(\bq_i) +\bOmega(\bq_1)+\frac{3}{2}\big(\bOmega(\bq_2)+\bOmega(\bq_3)\big)\big]\psi({-}\vbq_2)\psi({-}\vbq_3) + \ldots \nn
\end{align}
Extracting the component of appropriate dilatation weight, we therefore recover
\begin{align}
 \delta\<T^i_i(\vbq_1)\>_s  &=\bK^{-2}\bOmega_{(0)}(\bq_1)\psi_{(0)}(\vbq_1)  
 +\int[[\d\bq_2\d\bq_3]]\, \bK^{-2}\big[
{-}\bLambda_{(0)}(\bq_i)+\bOmega_{(0)}(\bq_1) \nn\\[1ex]& \qquad\qquad\qquad\qquad\qquad +\frac{3}{2}\big(\bOmega_{(0)}(\bq_2)+\bOmega_{(0)}(\bq_3)\big)\big]\psi_{(0)}({-}\vbq_2)\psi_{(0)}({-}\vbq_3) + \ldots, 
\end{align}
exactly as in the case of asymptotically AdS domain-walls.
The results \eqref{holo_result1} and \eqref{holo_result2} are thus valid for both asymptotically AdS and asymptotically power-law domain-walls.

\section{Holographic formulae for cosmology}
\label{Hol_formulae}

In Section \ref{Bispec_section}, we saw that the 
power spectrum and the bispectrum may be expressed in terms of the cosmological response functions, 
and in the previous section, we saw that the domain-wall response functions are related 
to 2- and 3-point functions of the dual QFT. We will now combine these results to obtain our main
holographic formulae for cosmology.

Combining the late-time cosmological result \eqref{cosmo_result2pt} with our holographic result \eqref{holo_result1},
we obtain the relation \cite{McFadden:2009fg, McFadden:2010na}
\[
\label{old_result}
\<\!\<\hat{\z}(q)\hat{\z}(-q)\>\!\> =-\frac{1}{2\Im\<\!\<T(-iq)T(iq)\>\!\>}.
\]
Using \eqref{cosmo_result} and our holographic result \eqref{holo_result2}, together with (\ref{old_result}),
we find
\begin{align}
\label{Main_result2}
&\<\!\<\hat{\z}(q_1)\hat{\z}(q_2)\hat{\z}(q_3)\>\!\> =-\frac{1}{4}\,\frac{1}{\prod_i \Im [\<\!\<T(-iq_i)T(iq_i)\>\!\>]}\,\Im \big[\<\!\<T(-iq_1)T(-iq_2)T(-iq_3)\>\!\> +\nn \\[1ex]
&\qquad \qquad  \sum_i \<\!\<T(-iq_i)T(iq_i)\>\!\> 
-2\big( \<\!\<T(-iq_1)\Upsilon(-iq_2,-iq_3)\>\!\>+\mathrm{cyclic\,perms}\big)\big].
\end{align}
This is our main result.  Using these formulae one may compute cosmological observables 
from QFT correlation functions.  These formulae were derived by working in the regime
where gravity is valid everywhere: we postulate however that they hold generally, as we will now explore.

\section{Holographic non-Gaussianity}
\label{HolNG_section}

Let us consider the case where the universe was non-geometric at early times,
with the dual QFT providing a perturbative description.  
At late times, there is 
an asymptotic expansion for the cosmological metric 
corresponding to the domain-wall asymptotic expansion \eqref{FG}.
(For QFTs of the form \eqref{Lfree}, this metric would be that of the dual frame; the late-time Einstein frame metric then takes the power-law form \eqref{power_law}).
Einstein's equations are therefore satisfied at late times, ensuring the conservation of $\z$ in the usual manner
and permitting a smooth transition to standard hot big bang cosmology.
To predict the spectrum and bispectrum of primordial scalar perturbations,
we simply need to work out the relevant QFT correlators and insert them in the 
holographic formulae above. The computation of the power spectrum is discussed in detail in
\cite{McFadden:2010na}. Here, we will compute the bispectrum.
We will work throughout in the large $\bN$ limit, with Euclidean signature metric.

\subsection{QFT results} 

\begin{figure}[t]
\center
\includegraphics[width=5.5cm]{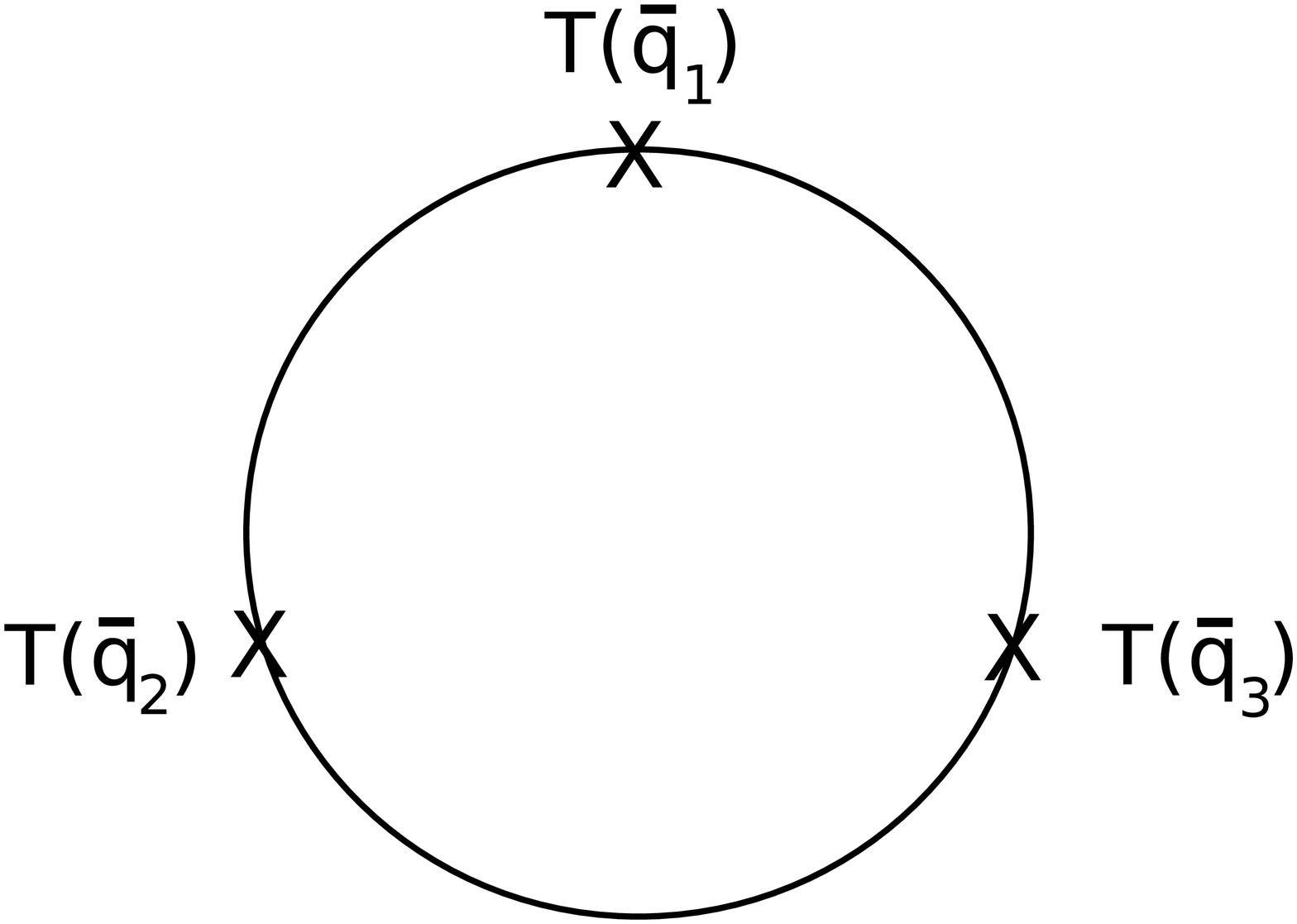}
\caption{\label{1-loopfig}
1-loop contribution to $\<T(\bq_1)T(\bq_2)T(\bq_3)\>$.
}
\end{figure}

From the QFT action \eqref{Lfree}, we see that propagators appear with a factor of $\gYM^2$, while vertices and insertions of the stress-energy tensor each contribute a factor of $1/\gYM^{2}$.  The leading contribution to the 3-point function $\<\!\<TTT\>\!\>$ comes therefore from the 1-loop diagram in Figure \ref{1-loopfig}, which is of order $\bN^2$ and involves only the free part of the Lagrangian.
Interactions contribute to diagrams at 2-loop order and higher, but these are suppressed by factors of $\geff^2$
relative to the 1-loop contribution and will be neglected here
(as discussed in \cite{McFadden:2009fg,McFadden:2010na}, $\geff^2$ is of the order of $n_s{-}1 \sim O(10^{-2})$).

For spatially flat cosmologies, the background metric seen by the dual QFT is also flat.  The dual stress-energy tensor is then given by
\[
 T_{ij} = T^A_{ij}+T^\phi_{ij}+T^\psi_{ij}+T^\chi_{ij},
\]
where the contributions from the various fields in \eqref{Lfree} are
\begin{align}
& T^A_{ij}=\frac{1}{g_{\mathrm{YM}}^2} \mathrm{tr} \big[2F^I_{ik}F^I_{jk}-\delta_{ij}\frac{1}{2}F^I_{kl}F^I_{kl}\big] + T^{\,\mathrm{ gauge-fixing}}_{ij}+T^{\,\mathrm{ghost}}_{ij}, \nn \\ 
& T^\phi_{ij} =\frac{1}{g_{\mathrm{YM}}^2} \mathrm{tr} \big[\D_i\phi^J \D_j\phi^J-\delta_{ij}\frac{1}{2}(\D\phi^J)^2\big], \nn \\
& T^\chi_{ij} = \frac{1}{g_{\mathrm{YM}}^2}\mathrm{tr} \big[\D_i\chi^K \D_j\chi^K-\frac{1}{8}\D_i\D_j(\chi^K)^2-\delta_{ij}\big(\frac{1}{2}(\D\chi^K)^2-\frac{1}{8}\D^2(\chi^K)^2\big)\big], \nn \\
& T^\psi_{ij} = \frac{1}{g_{\mathrm{YM}}^2}\mathrm{tr} \big[\frac{1}{2}\bar{\psi}^L\g_{(i}{\overleftrightarrow{\D}}_{j)}\psi^L - \delta_{ij}\half\bar{\psi}^L\overleftrightarrow{\slashed{\D}}\psi^L \big].
\end{align}
Here, we suppress the contribution to the stress-energy tensor from the interaction terms in (\ref{Lfree}),
as these terms do not contribute to the 1-loop computation.
Note that the trace of the stress-energy tensors for both conformally coupled scalars and for massless fermions vanish on shell.  This is a consequence of the Weyl invariance of the quadratic action for these fields (with the fields transforming nontrivially) when the action \eqref{Lfree} is appropriately coupled to gravity.

The 2-point function for the full stress-energy tensor $T_{ij}$ was evaluated in \cite{McFadden:2009fg, McFadden:2010na}.
Here, we need only the results
\begin{align}
\label{TTresults}
&\frac{1}{\mathcal{N}_A}\<\!\<T^A(\bq)T^A(-\bq)\>\!\> = \frac{1}{\mathcal{N}_\phi}\<\!\<T^\phi(\bq)T^\phi(-\bq)\>\!\> = \frac{1}{64}\bN^2\bq^3 + O(\geff^2),
\nn \\[1ex]
&\qquad\qquad \<\!\<T^\psi(\bq)T^\psi(-\bq)\>\!\> = \<\!\<T^\chi(\bq)T^\chi(-\bq)\>\!\> = O(\geff^2).
\end{align}
As for the 3-point function, the 1-loop contribution from minimally coupled scalars shown in Figure \ref{1-loopfig} is given by
\begin{align}
\label{TTTphi}
\<\!\<T^\phi(\bq_1)T^\phi(\bq_2)T^\phi(\bq_3)\>\!\> &= -\mathcal{N}_\phi \bN^2\int[\d\bq] \,\frac{\bq{\cdot}(\bq+\bq_1)\, \bq{\cdot}(\bq-\bq_2)\, (\bq+\bq_1){\cdot}(\bq-\bq_2)}{\bq^2(\bq+\bq_1)^2(\bq-\bq_2)^2} + O(\geff^2) \nn \\[1ex]
&=\frac{1}{128}\mathcal{N}_\phi\bN^2\big(2\bq_1\bq_2\bq_3-(\bq_1+\bq_2+\bq_3)(\bq_1^2+\bq_2^2+\bq_3^2)\big) + O(\geff^2),
\end{align}
where in the evaluation of the integral we used the result\footnote{Note the l.h.s.~reduces to a standard integral upon inverting all momenta, $\vbq_i{}'=\vbq_i/\bq_i^2$.}
\[
 \int[\d\bq]\,\, \frac{1}{\bq^2(\bq+\bq_1)^2(\bq-\bq_2)^2} = \frac{1}{8\, \bq_1\bq_2\bq_3}.
\]
For gauge fields, the corresponding 1-loop contribution is given by
\[
 \<\!\<T^A(\bq_1)T^A(\bq_2)T^A(\bq_3)\>\!\> = \mathcal{N}_A \bN^2 \int [\d\bq]\, \pi_{ij}(\bq)\pi_{jk}(\bq+\bq_1)\pi_{ki}(\bq-\bq_2) +O(\geff^2),
\]
where the projection operator $\pi_{ij}(\bq)=\delta_{ij}-\bq_i\bq_j/\bq^2$ and the contributions from the ghost and gauge-fixing terms cancel out, as one may have anticipated since the contribution of these terms to the 
stress-energy tensor is BRST-exact.
Evaluating this integral, we find
\[
\label{TTTA}
\<\!\<T^A(\bq_1)T^A(\bq_2)T^A(\bq_3)\>\!\> =\frac{\mathcal{N}_A}{\mathcal{N}_\phi} \Big[\<\!\<T^\phi(\bq_1)T^\phi(\bq_2)T^\phi(\bq_3)\>\!\> +2\sum_i\<\!\<T^\phi(\bq_i)T^\phi(-\bq_i)\>\!\>\Big] +O(\geff^2).
\]
The fact that the 3-point function of the vector fields is related to that of the scalars is not unexpected, since the
vector fields are dual to scalar fields in three dimensions.

Massless fermions and conformally coupled scalars, however, make no contribution to the 3-point function:
\[
 \<\!\<T^\chi(\bq_1)T^\chi(\bq_2)T^\chi(\bq_3)\>\!\>=\<\!\<T^\psi(\bq_1)T^\psi(\bq_2)T^\psi(\bq_3)\>\!\>=O(\geff^2).
\]

The $\<\!\<T\Upsilon\>\!\>$ terms may be determined by direct calculation.  In position space,
\begin{align}
 & \Upsilon^\phi(\x_1,\x_2) = 0, \qquad \Upsilon^A(\x_1,\x_2) = T^A(\x_1) \delta(\x_1-\x_2), \nn \\
 & \Upsilon^\chi(\x_1,\x_2) = -\frac{1}{8}\frac{\partial}{\partial x_1^i}\big[(\chi^K(\x_1))^2\frac{\partial}{\partial x_1^i}\delta(\x_1-\x_2)\big], \nn \\
& \Upsilon^\psi(\x_1,\x_2) = -\frac{1}{2}T^\psi(\x_1)\delta(\x_1-\x_2)-\bar{\psi}^L(\x_1)\gamma^i\psi^L(\x_1)\frac{\partial}{\partial x_1^i}\delta(\x_1-\x_2).
\end{align}
We then find
\begin{align}
\label{TY}
 & \<\!\<T^\phi(\bq_1)\Upsilon^\phi(\bq_2,\bq_3)\>\!\> =0, \quad  \<\!\<T^\chi(\bq_1)\Upsilon^\chi(\bq_2,\bq_3)\>\!\> =O(\geff^2), \quad
 \<\!\<T^\psi(\bq_1)\Upsilon^\psi(\bq_2,\bq_3)\>\!\> =O(\geff^2), \nn\\
 & \qquad\qquad \<\!\<T^A(\bq_1)\Upsilon^A(\bq_2,\bq_3)\>\!\> = \<\!\<T^A(\bq_1)T^A(\bq_1)\>\!\> =\frac{\mathcal{N}_A}{\mathcal{N}_\phi}\,\<\!\<T^\phi(\bq_1)T^\phi(-\bq_1)\>\!\>.
\end{align}

Putting everything together, in the denominator of \eqref{Main_result2}, we have
\[
\<\!\<T(\bq)T(-\bq)\>\!\> = \frac{1}{64}(\mathcal{N}_A+\mathcal{N}_\phi)\bN^2\bq^3 +O(\geff^2),
\]
while in the numerator,  
\begin{align}
&\<\!\<T(\bq_1)T(\bq_2)T(\bq_3)\>\!\> + \sum_i \<\!\<T(\bq_i)T(-\bq_i)\>\!\>-2\big(\<\!\<T(\bq_1)\Upsilon(\bq_2,\bq_3)\>\!\> +\mathrm{cyclic\,\,perms}\big)\nn\\
&\qquad\qquad\qquad =\frac{1}{128}(\mathcal{N}_A+\mathcal{N}_\phi)\bN^2\big(2\bq_1\bq_2\bq_3+\sum_i \bq_i^3-(\bq_1\bq_2^2+5\,\mathrm{perms})\big)+O(\geff^2).
\end{align}
Note that the 2-point terms on the r.h.s.~of \eqref{TTTA} cancel the $\<\!\<T^A\Upsilon^A\>\!\>$ term in \eqref{TY} leaving a result in which the dependence on the number of fields appears only as an overall prefactor of $(\mathcal{N}_A+\mathcal{N}_\phi)$.

\subsection{Holographic prediction for the bispectrum} 

Analytically continuing $\bN$ and $\bq$ according to \eqref{QFT_cont}, the holographic formulae \eqref{old_result} and \eqref{Main_result2} yield, to leading order in $\geff^2$, the results
\begin{align}
\label{bispec}
\<\!\<\hat{\z}(q)\hat{\z}(-q)\>\!\> &= \frac{32}{\mathcal{N}N^2q^3}, \nn\\[1ex]
\<\!\<\hat{\z}(q_1)\hat{\z}(q_2)\hat{\z}(q_3)\>\!\>
&= \frac{512}{\mathcal{N}^2 N^4} \Big(\prod_i q_i^{-3}\Big)\big({-}2q_1q_2q_3-\sum_i q_i^3+(q_1q_2^2+5\,\mathrm{perms})\big),
\end{align}
where $\mathcal{N}=\mathcal{N}_A+\mathcal{N}_\phi$.
(Note that the power spectrum is no longer exactly scale invariant when we include $O(\geff^2)$ corrections, as detailed in \cite{McFadden:2009fg, McFadden:2010na}).

Interestingly, these results are an {\it exact} fit to the factorisable equilateral template introduced in \cite{Creminelli:2005hu}, for which
\[
\label{template}
\<\!\<\hat{\z}(q_1)\hat{\z}(q_2)\hat{\z}(q_3)\>\!\> =
6A^2 \fnleq\Big(\frac{3}{5}\Big) \Big(\prod_i q_i^{-3}\Big)\big({-}2q_1q_2q_3-\sum_i q_i^3+(q_1q_2^2+5\,\mathrm{perms})\big),
\]
where $A = q^3\<\!\<\hat{\z}(q)\hat{\z}(-q)\>\!\>$.
Comparing with \eqref{bispec}, we see that $A = 32/\mathcal{N}N^2$ and
\[
\label{fnlresult}
 \fnleq = 5/36.
\]
Remarkably, this result is completely independent of the field content of the dual QFT.
The fact that we have only the equilateral type non-Gaussianity present, and not the local type, stems from the presence of the 2-point terms in our holographic formula \eqref{Main_result2}.  These serve to cancel out the local-type non-Gaussianity present in QFT 3-point correlators such as \eqref{TTTphi}, yielding a cosmological bispectrum of the purely equilateral form.

Note that in \eqref{fnlresult} we are using the WMAP sign convention for $\fnleq$.  The template \eqref{template} follows from that given in \cite{Komatsu:2010hc} for $\Phi_B$, the Bardeen curvature potential in the matter-dominated era, using the linearised relation $\Phi_B=(3/5)\z$.
Other authors have preferred to define $\fnl$ using the Newtonian potential in place of $\Phi_B$, which yields a value for $\fnl$ with the opposite sign.  Physically, the two conventions are easily distinguished: with the WMAP convention used here, a positive value for $\fnl$ implies an excess of cold spots in the CMB.

\section{Discussion}

In this paper, we presented a holographic description of inflationary cosmology at quadratic order in perturbation theory, and initiated the holographic analysis of cosmological non-Gaussianity.
Our most important result is the holographic formula \eqref{Main_result2} expressing the cosmological bispectrum of curvature perturbations in terms of correlation functions of the dual QFT.  With the aid of this formula, we have computed the non-Gaussianity of cosmological perturbations at the end of a primordial holographic inflationary epoch in which gravity was strongly coupled at very early times.  The resulting non-Gaussianity is of {\it exactly} the factorisable equilateral type with $\fnleq=5/36$, irrespective of all details of the dual QFT.  It is striking that a simple holographic phenomenological approach yields such clear, unambiguous predictions.

While an $\fnl$ of this magnitude is almost certainly too small to be directly measurable with the Planck satellite, the detection of much larger $\fnl$ values would of course eliminate the present class of holographic models.  (Here, it should be understood that our prediction for $\fnl$ refers specifically to the residual primordial component once all non-Gaussianities arising from the post-inflationary evolution have been subtracted out).

The holographic models explored here may be distinguished from their conventional inflationary counterparts through a combination of the predicted running of the spectral index discussed in \cite{McFadden:2010jw}, along with the predictions for the scalar bispectrum discussed above.  Conventional inflationary models (meaning those based on weakly coupled gravity) that predict non-Gaussianity of the equilateral type include: ghost inflation \cite{ArkaniHamed:2003uz} and DBI inflation \cite{Alishahiha:2004eh}, which typically predict $\fnleq \sim 100$; tilted ghost inflation where $\fnleq \gtrsim 1$ \cite{Senatore:2004rj}; and slow-roll inflation with higher derivative couplings where $\fnleq \lesssim 1$ \cite{Creminelli:2003iq}.  Nevertheless, in all these cases, the bispectrum is only approximately of the equilateral type, unlike the exact equilateral form found in the case of the holographic models.

Finally, let us remark that the methods we have developed in this paper are straightforwardly applicable to other forms of holographic non-Gaussianity.  We will report holographic results for cosmological 3-point correlators involving tensors in a forthcoming publication \cite{McFadden:2011kk}.  Beyond this, the holographic calculation of the scalar trispectrum\footnote{\renewcommand{\baselinestretch}{1}\normalsize\small
Parametrically, we expect $\<\z\z\z\z\>\sim \<TTTT\>/\<TT\>^4\sim N^{-6}$ and so the nonlinearity parameter $\gnl$ defined by $\<\z\z\z\z\>\sim \gnl \<\z\z\>^3$ ought to be independent of $N$ at leading order in the large-$N$ limit.
The same is expected to hold for all similarly defined higher order nonlinearity
parameters.
} appears a worthy target for future endeavours.

\vskip 0.3cm

{\it Acknowledgments:} We thank Adam Bzowski for discussions. 
The authors are supported by NWO, the Nederlandse Organisatie voor Wetenschappelijk Onderzoek.


\appendix


\section{Gauge-invariant variables at second order} 
\label{App_GT}

Decomposing the metric to second order as $g_{\mu\nu} = g^{(0)}_{\mu\nu}+\delta g_{\mu\nu}$, the metric perturbation $\delta g_{\mu\nu}$ transforms under a second-order gauge transformation $\xi^\mu$ as 
\[
\label{myGT}
 \delta\check{g}_{\mu\nu} = \delta g_{\mu\nu}+\Lie_\xi g^{(0)}_{\mu\nu}+\Lie_\xi \delta g_{\mu\nu}+
 \frac{1}{2} \Lie^2_\xi g^{(0)}_{\mu\nu}.
\]
Readers familiar with \cite{Bruni:1996im, Matarrese:1997ay} may wish to verify that upon setting
\[
\label{GTexpansion}
\delta g_{\mu\nu}=\lambda \delta g_{\mu\nu}^{(1)}+\frac{\lambda^2}{2}\, \delta g_{\mu\nu}^{(2)}+O(\lambda^3), \qquad \xi^\mu = \lambda \xi_{(1)}^\mu+\frac{\lambda^2}{2}\, \xi_{(2)}^\mu + O(\lambda^3),
\]
and expanding in powers of $\lambda$, we recover the equivalent expressions
\[
\label{BruniGT}
\delta\check{g}_{\mu\nu}^{(1)} = \delta g_{\mu\nu}^{(1)}+\Lie_{\xi_{(1)}}g_{\mu\nu}^{(0)}, \qquad
\delta \check{g}_{\mu\nu}^{(2)} = \delta g_{\mu\nu}^{(2)}+\Lie_{\xi_{(2)}}g_{\mu\nu}^{(0)}
+\Lie^2_{\xi_{(1)}}g_{\mu\nu}^{(0)}+2\Lie_{\xi_{(1)}}\delta g_{\mu\nu}^{(1)}.
\]
In the present paper we do not write this expansion in $\lambda$ explicitly, but rather we simply work to quadratic order in the overall perturbation $\delta g_{\mu\nu}$.
We will therefore use \eqref{myGT} rather than \eqref{BruniGT} in the following.  (If desired, though, the expansion in $\lambda$ may be re-instated at any time using \eqref{GTexpansion}).

The transformed metric perturbations, decomposed as in \eqref{metric_perts}, are then
\begin{align}
\label{transformed_perts}
&\check{\phi} = (1/2)\sigma\delta\check{g}_{00},  \qquad
&\check{\nu}_i = a^{-2}\pi_{ij}\delta\check{g}_{0j},  \nn \\
&\check{\nu} = a^{-2}\D^{-2}\D_i \delta\check{g}_{0i}, \qquad
&\check{\omega}_i = a^{-2}\pi_{ij}\D_k \D^{-2}\delta\check{g}_{jk}, \nn\\
&\check{\psi} = -(1/4)a^{-2}\pi_{ij}\delta\check{g}_{ij}, \qquad
&\check{\g}_{ij} = a^{-2} \Pi_{ijkl}\delta\check{g}_{kl}, \nn\\
&\check{\chi} = (1/2)a^{-2}(\delta_{ij}-(3/2)\pi_{ij})\D^{-2}\delta\check{g}_{ij},
\end{align}
where the transverse and transverse traceless projection operators are
\[
 \pi_{ij}=\delta_{ij}-\D_i\D_j/\D^2, \qquad \Pi_{ijkl} = (1/2)(\pi_{ik}\pi_{jl}+\pi_{il}\pi_{jk}-\pi_{ij}\pi_{kl}).
\]

These formulae may then be evaluated explicitly as required.  Writing $\xi^\mu = (\alpha, \delta^{ij}\xi_j)$ (where $\xi_i$ may be further decomposed as $\xi_i = \beta_{,i}+\g_i$, where $\g_i$ is transverse), we find that, for example,
\bea
\label{psi_transf}
 \check{\psi} &=& \psi - H\alpha -\big(\frac{\dot{H}}{2}+H^2\big)\alpha^2-\frac{H}{2}\alpha\dot{\alpha}-\frac{H}{2}\xi_i\alpha_{,i} \nn \\
&& -\frac{1}{4}\pi_{ij}\Big( \alpha\dot{h}_{ij}+2H\alpha h_{ij}+\xi_k h_{ij,k}+\frac{2}{a^2} \dN_i\alpha_{,j}+2\xi_{k,i}h_{jk}+\frac{\sigma}{a^2} \alpha_{,i}\alpha_{,j}
+\xi_{k,i}\xi_{k,j}+4H\alpha\xi_{i,j}\Big). \nn \\
\eea
Similarly, the scalar field perturbation transforms as
\bea
\label{scalar_transf}
 \delta\check{\vphi} &=& \delta\vphi +\Lie_\xi \vphi +\Lie_\xi \delta \vphi+(1/2) \Lie^2_\xi \vphi \nn \\
&=& \delta\vphi + \alpha\dot{\vphi}+\alpha\delta\dot{\vphi}+\xi_i\delta\vphi_{,i}+(1/2)\ddot{\vphi}\alpha^2+(1/2)\dot{\vphi}\alpha\dot{\alpha}+(1/2)\dot{\vphi}\xi_i\alpha_{,i}.
\eea

To identify the gauge-invariant definition of $\z$, we consider transforming from a general gauge to the fully-fixed comoving gauge where
\[
\label{zeta_gauge}
 g^{co}_{ij} = a^2 e^{2\z}[e^{\hat{\g}}]_{ij} = a^2 [ \delta_{ij}+(2\z\delta_{ij}+\hat{\g}_{ij})+
(2\z^2\delta_{ij}+2\z\hat{\g}_{ij}+\half\hat{\g}_{ik}\hat{\g}_{kj})], \quad \delta\vphi^{co}=0.
\]
Recalling that $\hat{\g}_{ij}$ is transverse traceless, we see that $\chi^{co}$ and $\omega^{co}_i$ vanish at linear order 
in perturbation theory.  Since at linear order
\[
 \chi^{co} = \chi+\beta, \qquad \omega^{co}_i=\omega_i+\g_i,
\]
we therefore require $\alpha = -\delta\vphi/\dot{\vphi}$ and $\xi_i =-\hat{\xi}_i$ at linear order, where $\hat{\xi}_i\equiv \chi_{,i}+\omega_i$.
Using these first order quantities, we may then solve \eqref{scalar_transf} to set $\delta\vphi^{co}=0$ at quadratic order.  To pass to comoving gauge at quadratic order thus requires a change of slicing
\[
 \alpha = -\frac{\delta\vphi}{\dot{\vphi}}+\frac{\delta\vphi\delta\dot{\vphi}}{2\dot{\vphi}^2}+\frac{\hat{\xi}_i\delta\vphi_{,i}}{2\dot{\vphi}}.
\]
(One may similarly solve for $\xi_i$ to second order, however we will not need this quantity here).

Substituting these results into \eqref{psi_transf} yields\footnote{This result agrees with \cite{Malik:2003mv}, up to the spatial gradient terms that are dropped in this reference.}
\bea
\psi^{co} &=& \psi+\frac{H}{\dot{\vphi}}\delta\vphi -\frac{H}{\dot{\vphi}^2}\delta\vphi\delta\dot{\vphi}-\frac{H}{\dot{\vphi}}\hat{\xi}_k\delta\vphi_{,k}+\Big(\frac{H\ddot{\vphi}}{2\dot{\vphi}}-\frac{\dot{H}}{2}-H^2\Big)\frac{\delta\vphi^2}{\dot{\vphi}^2} \nn \\
&& +\frac{1}{4}\pi_{ij}\Big(\frac{2}{a^2\dot{\vphi}}\dN_i\delta\vphi_{,j}+\frac{2H}{\dot{\vphi}}\delta\vphi h_{ij}+\frac{\delta\vphi}{\dot{\vphi}}\dot{h}_{ij} + 2\hat{\xi}_{k,i}h_{jk}+\hat{\xi}_k h_{ij,k} \nn \\
&& \qquad\qquad
-\frac{\sigma}{a^2\dot{\vphi}^2}\delta\vphi_{,i}\delta\vphi_{,j}-\frac{4H}{\dot{\vphi}}\delta\vphi \hat{\xi}_{i,j}-\hat{\xi}_{k,i}\hat{\xi}_{k,j}
\Big).
\eea
On the other hand, from \eqref{transformed_perts} and \eqref{zeta_gauge}, 
\[
 \psi^{co} = -\z-\z^2-\frac{1}{4}\pi_{ij}(2\z\hat{\g}_{ij}+\half\hat{\g}_{ik}\hat{\g}_{kj}).
\]
Upon inversion, this yields
\[
 \z = -\psi^{co}-(\psi^{co})^2+\frac{1}{4}\pi_{ij}(2\psi^{co}\g_{ij}-\half\g_{ik}\g_{kj}),
\]
using the fact that $\hat{\g}_{ij}=\g_{ij}$ to linear order.  
The gauge-invariant expression \eqref{zeta_gi} for $\z$ at quadratic order then follows directly.
We have checked the gauge-invariance of this expression explicitly.

\section{Constraint equations}

\label{App_constraints}

In this appendix, we present the domain-wall Hamiltonian and momentum constraint equations to quadratic order, as required for our holographic calculations in Section \ref{TTT_section}.  We will assume the shift has been gauged to zero ($N_i=0$), but otherwise retain all perturbations.

The full Hamiltonian constraint reads
\[
 0=-R + K^2 -K_{ij}K^{ij}+2\bK^2 V -N^{-2}\dot{\Phi}^2+g^{ij}\Phi_{,i}\Phi_{,j},
\]
where $K_{ij}=(1/2N)\dot{g}_{ij}$ is the extrinsic curvature of constant-$z$ slices.
Expanding to quadratic order, we find
\bea
\label{Ham_constr}
 0 &=& -4a^{-2}\D^2\psi+2H\dot{h}+4\bK^2 V \dN-2\dot{\vphi}\delta\dot{\vphi}+2\bK^2 V' \delta\vphi \nn \\
&& -R_{(2)} + \frac{1}{4}\dot{h}^2-\frac{1}{4}\dot{h}_{ij}\dot{h}_{ij}-2Hh_{ij}\dot{h}_{ij}-4H\dot{h}\,\dN-6\bK^2 V \dN^2 \nn \\
&& -\delta\dot{\vphi}^2+4\dot{\vphi}\,\dN\delta\dot{\vphi}+\bK^2 V'' \delta\vphi^2+a^{-2}\delta\vphi_{,i}\delta\vphi_{,i},
\eea
where repeated covariant indices are to be summed over using the Kronecker delta, and $h \equiv h_{ii}$.
For the purposes of our holographic calculations, we will not need to evaluate $R_{(2)}$ explicitly.

Similarly, the momentum constraint
\[
 0 = \nabla_j (K^j_i -\delta^j_i K)-N^{-1}\dot{\Phi}\Phi_{,i},
\]
yields
\bea
 0 &=& \half\dot{h}_{ij,j}-\half\dot{h}_{,i}+2H\dN_{,i}-\dot{\vphi}\delta\vphi_{,i} \nn \\
&& +\frac{1}{4}h_{,j}\dot{h}_{ji}-\frac{1}{4}\dot{h}_{jk} h_{jk,i} -\delta\dot{\vphi}\delta\vphi_{,i}+\dot{\vphi}\,\dN\delta\vphi_{,i}-4H\dN\dN_{,i} \nn \\
&& +\half (h_{jk}\dot{h}_{jk}+\dot{h}\,\dN)_{,i}-\half (h_{jk}\dot{h}_{ki} +\dot{h}_{ij}\,\dN)_{,j}
\eea
when expanded to quadratic order.
Extracting the scalar part by acting with $\D^{-2}\D_i$, we find
\bea
\label{mom_constr}
 0 &=& 2\dot{\psi}-\dot{\vphi}\delta\vphi+2H\dN \nn \\
&&+ \half h_{jk}\dot{h}_{jk}-2H\dN^2+\half\dot{h}\,\dN \nn \\
&& +\D^{-2}\D_i\Big[ \frac{1}{4}h_{,j}\dot{h}_{ji}-\frac{1}{4}\dot{h}_{jk}h_{jk,i}-\half (h_{jk}\dot{h}_{ki}+\dot{h}_{ij}\dN)_{,j}-\delta\dot{\vphi}\delta\vphi_{,i}+\dot{\vphi}\,\dN\delta\vphi_{,i}\Big].
\eea


\end{document}